\documentclass[12pt,journal,onecolumn,draftcls]{IEEEtran}
\usepackage{epsfig}
\usepackage{times}
\usepackage{float}
\usepackage{afterpage}
\usepackage{amsmath}
\usepackage{amstext}
\usepackage{amssymb,bm}
\usepackage{latexsym}
\usepackage{color}
\usepackage{graphicx}
\usepackage{amsmath}
\usepackage{amsthm}
\usepackage{graphicx}
\usepackage[center]{caption}
\usepackage{pstricks}
\usepackage{subfigure}
\usepackage{booktabs}
\usepackage{multicol}
\usepackage{lipsum}
\usepackage{dblfloatfix}
\usepackage{algpseudocode}
\usepackage[ruled,vlined]{algorithm2e}

\allowdisplaybreaks

\def \eu{\mathrm{e}}
\def \jj{\mathrm{j}}

\newcommand{\Tr}{\mathrm{Tr}}

\newtheorem{thm}{Theorem}

\newtheorem{prop}[thm]{Proposition}
\newtheorem{rem}{Remark}
\newtheorem{defin}{Definition}

\begin{document}

\title{On the Optimality of {\it Simple} Schedules for Networks with Multiple Half-Duplex Relays}

\author{Martina~Cardone, Daniela~Tuninetti and Raymond~Knopp
\thanks{
M.~Cardone and R.~Knopp are with the Mobile Communications Department at Eurecom, Biot, 06410, France (e-mail: cardone@eurecom.fr; knopp@eurecom.fr). 
Eurecom's research is partially supported by its industrial partners: BMW Group Research \& Technology, IABG, Monaco Telecom, Orange, SAP, SFR, ST Microelectronics, Swisscom and Symantec. The research at Eurecom leading to these results has received funding from the EU Celtic+ Framework Program Project SHARING and from a 2014 Qualcomm Innovation Fellowship.

D. Tuninetti is with the Electrical and Computer Engineering Department of the University of Illinois at Chicago, Chicago, IL 60607 USA (e-mail: danielat@uic.edu). 
The work of D.~Tuninetti was partially funded by NSF under award number 1218635;
the contents of this article are solely the responsibility of the author and
do not necessarily represent the official views of the NSF.
D.~Tuninetti would like to acknowledge insightful discussions with Dr. Salim~El~Rouayheb on sumbodular functions.

The results in this paper were submitted in part to the 2015 IEEE Information Theory Workshop. 
}

}\maketitle
\begin{abstract}
This paper studies networks with $N$ half-duplex relays assisting the communication between a source and a destination. In ISIT'12 Brahma, \"{O}zg\"{u}r and Fragouli conjectured that in Gaussian half-duplex diamond networks (i.e., without a direct link between the source and the destination, and with $N$ non-interfering relays) an approximately optimal relay scheduling policy (i.e., achieving the cut-set upper bound to within a constant gap) has at most $N+1$ active states (i.e., at most $N+1$ out of the $2^N$ possible relay listen-transmit states have a strictly positive probability). Such relay scheduling policies were referred to as simple. In ITW'13 we conjectured that simple approximately optimal relay scheduling policies exist for any Gaussian half-duplex multi-relay network irrespectively of the topology. This paper formally proves this more general version of the conjecture and shows it holds beyond Gaussian noise networks. In particular, for any memoryless half-duplex $N$-relay network with independent noises and for which independent inputs are approximately optimal in the cut-set upper bound, an approximately optimal simple relay scheduling policy exists. A convergent iterative polynomial-time algorithm, which alternates between minimizing a submodular function and maximizing a linear program, is proposed to find the approximately optimal simple relay schedule. As an example, for $N$-relay Gaussian networks with independent noises, where each node in equipped with multiple antennas and where each antenna can be configured to listen or transmit irrespectively of the others, the existence of an approximately optimal simple relay scheduling policy with at most $N+1$ active states is proved. Through a line-network example it is also shown that independently switching the antennas at each relay can provide a strictly larger multiplexing gain compared to using the antennas for the same purpose.
\end{abstract}


\section{Introduction}
Adding relaying stations to today's cellular infrastructure promises to boost network performance in terms of coverage, network throughput and robustness. Relay nodes, in fact, provide extended coverages in targeted areas, offering a way through which the base station can communicate with cell-edge users. Moreover, the use of relay nodes may offer a cheaper and lower energy consumption alternative to installing new base stations, especially for regions where deployment of fiber fronthaul solutions are impossible. Depending on the mode of operation, relays are classified into two categories: Full-Duplex (FD) and Half-Duplex (HD). A relay is said to operate in FD mode if it can receive and transmit simultaneously over the same time-frequency-space resource, and in HD mode otherwise. Although higher rates can be attained with FD relays, due to practical restrictions (such as the inability to perfectly cancel the self-interference \cite{Duarte,Everett}) currently employed relays operate in HD mode, unless sufficient isolation between the antennas can be achieved.

Motivated by the current practical importance of relaying stations, in this paper we study networks where the communication between a source and a destination is assisted by $N$ HD relays. In particular each relay is assumed to operate in time division duplexing, i.e., in time it alternates between transmitting and receiving. In such a network 
there are $2^N$ possible listen-transmit states whose probability must be optimized. 
Due to the prohibitively large complexity of this optimization problem (i.e., exponential in the number of relays $N$) it is critical to identify, if any, structural properties of such networks that can be leveraged in order to find optimal solutions with limited complexity.
This paper uses properties of submodular functions and Linear Programs (LP) to show that a class of memoryless HD multi-relay networks has indeed intrinsic structural properties that guarantee the existence of approximately optimal simple relay scheduling policies that can be determined in polynomial time.

\subsection{Related Work}
The different relaying strategies studied in the literature are largely based on the seminal work by Cover and El Gamal~\cite{coveElGamal} on memoryless FD relay channels. In~\cite{coveElGamal} the authors proposed a general outer bound (now known as the max-flow min-cut outer bound, or cut-set for short) and two achievable strategies named Decode-and-Forward (DF) and Compress-and-Forward (CF). 
In~\cite{KramerGastparGuptaIT05}, these bounds were extended to networks with multiple FD relays. 
The capacity of a multi-relay network is open in general. 
In~\cite{avestimher:netflow}, the authors showed that for Gaussian noise networks with $N$ FD relays Quantize-reMap-and-Forward (QMF)---a network generalization of CF---achieves the cut-set upper bound to within $\sum_{k=1}^{N+2} 5 \min\{M_k,N_k\}$ bits per channel use, with $M_k$ and $N_k$ being the number of transmit and receive antennas, respectively, of node $k\in[1:N+2]$.
For single-antenna nodes, this gap was reduced to $1.26(N + 2)$ bits per channel use in~\cite{nncLim} by means of a novel transmission strategy named Noisy Network Coding (NNC)---also a network generalization of CF. 
%
In \cite{arXiv:1207.5660,SenguptaITW2012}, the authors showed that for Gaussian FD multi-relay networks with a sparse topology, namely {\it diamond} networks without a direct source-destination link and with $N$ FD non-interfering relays, the gap is of $2\log(N+1)$ bits per channel use.

Relevant past work on HD multi-relay networks comprises the following papers.
By following the approach of~\cite{kramer-allerton}, in~\cite{Ozgur2010} the authors evaluated the cut-set upper bound for Gaussian multi-relay networks and, for the case of single-antenna nodes, they showed that a lattice-code implementation of QMF is optimal to within $8(N + 2)$ bits per channel use \cite[Theorem 2.3]{Ozgur2010}.
Recently, in~\cite{ourITjournal} we showed that the gap can be reduced to $1.96\left(N+2\right)$ bits per channel use by using NNC.
In general, finding the capacity of a single-antenna Gaussian HD multi-relay network is a combinatorial problem since the cut-set upper bound is the minimum between $2^N$ bounds (one for each possible cut in the network), each of which is a linear combination of $2^N$ relay states (since each relay can either transmit or receive). Thus, as the number of relays increases, optimizing the cut-set bound becomes prohibitively complex. Identifying structural properties of the cut-set upper bound, or of a constant gap approximation of the cut-set upper bound, is therefore critical for efficient numerical evaluations and can have important practical consequences for the design of {\it simple} / reduced complexity relay scheduling policies. 

In~\cite{Bagheri2009}, the authors analyzed the single-antenna Gaussian HD diamond network with $N=2$ relays and proved that at most $N+1=3$ states, out of the $2^N=4$ possible ones, suffice to approximately (to within a constant gap) characterize the capacity.
We say that 
these $N+1$ states are {\it active} (have a strictly positive probability) and form an (approximately) optimal {\it simple} schedule.
In~\cite{Fragouli2012}, Brahma {\it et al} verified through extensive numerical evaluations that single-antenna Gaussian HD diamond networks with $N \leq 7$ relays have (approximately) optimal simple schedules and conjectured this to be true for any $N$. 
In~\cite{BrahmaISIT2014}, Brahma {\it et al}'s conjecture was proved for single-antenna Gaussian HD diamond networks with $N \leq 6$ relays; the proof is by contradiction 
and uses properties of submodular functions and LP duality but requires numerical evaluations; for this reason the authors could only prove the conjecture for $N \leq 6$, since for larger values of $N$ ``the computational burden becomes prohibitive''~\cite[page 1]{BrahmaISIT2014}.
Our numerical experiments in~\cite{ourITW2014} showed that Brahma {\it et al}'s conjecture 
holds for general single-antenna Gaussian HD multi-relay networks (i.e., not necessarily with a diamond topology) with $N \leq 8$; we conjectured that the same holds for any $N$. 
If our more general version of Brahma {\it et al}'s conjecture is true, then single-antenna Gaussian HD multi-relay networks have (approximately) optimal simple schedules irrespectively of their topology, i.e., known results for diamond networks are not a consequence of the simplified / sparse network topology. 
In this work, we formally prove the conjecture for a general  Multiple-Input-Multiple-Output (MIMO) Gaussian HD multi-relay network
and show that this result holds beyond Gaussian noise networks.

In~\cite{ourITjournal} we also discussed polynomial-time algorithms to determine the (approximately) optimal simple schedule and their extensions beyond relay networks. Other algorithms seeking to determine optimal relay scheduling policies, but not focused on characterizing the minimum number of active states, are available in the literature.
The authors of~\cite{OngMultiRelay} proposed an iterative algorithm to determine the optimal schedule when the relays use DF.
In~\cite{EtkinParvareshShomoronyAvestimehr} the authors proposed a `grouping' technique to find the relay schedule that maximizes the approximate capacity of certain Gaussian HD relay networks, including for example layered networks; 
because finding a good node grouping is computationally complex, the authors proposed an heuristic approach based on tree decomposition that results in polynomial-time algorithms; as for diamond networks in~\cite{Fragouli2012}, 
the low-complexity algorithm of~\cite{EtkinParvareshShomoronyAvestimehr} relies on the `simplified' topology of certain networks.
As opposed to these works, we propose a polynomial-time algorithm that determines the (approximately) optimal simple relay policy with a number of active states at most equal to the number of relays plus one for any network topology.

The first step in the derivation of our main result uses~\cite[Theorem 1]{ParvaEtk} that states that for FD relay networks ``under the assumption of independent inputs and noises, the cut-set bound is submodular''; wireless erasure networks, Gaussian networks and their linear deterministic high-SNR approximations are examples for which~\cite[Theorem 1]{ParvaEtk} holds.

\subsection{Contributions}
In this work we study multi-relay HD networks. In particular, we seek to identify 
properties of the network that allow for the reduction of the complexity in computing an (approximately) optimal relay scheduling policy. 
Our main contributions can be summarized as follows:
\begin{enumerate}
\item 
We formally prove Brahma {\it et al}'s conjecture beyond the Gaussian noise case. In particular, we prove that for any HD network with $N$ relays, with independent noises and for which independent inputs in the cut-set bound are approximately optimal, the optimal relay policy is simple. The key idea is to use the Lov\'{a}sz extension and the greedy algorithm for submodular polyhedra to highlight structural properties of the minimum of a submodular function. Then, by using the saddle-point property of min-max problems and the existence of optimal basic feasible solutions for LPs, an (approximately) optimal relay policy with the claimed number of active states can be shown.
\item 
We propose an iterative algorithm to find the (approximately) optimal simple relay schedule, which alternates between minimizing a submodular function and maximizing a LP. The algorithm runs in polynomial-time (in the number of relays $N$) since the unconstrained minimization of a submodular function can be performed in strongly polynomial-time and a LP maximization can also be performed in polynomial-time.
\item 
For Gaussian noise networks with multi-antenna nodes, where the antennas at the relays may be switched between transmit and receive modes independently of one another, we prove that NNC is optimal to within $1.96$~bits per channel use per antenna, and that an (approximately) optimal schedule has at most $N+1$ active states (as in the single-antenna case) regardless of the total number of antennas in the system. We also show, through two examples, that switching independently the antennas at each relay achieves in general higher rates than using all of them for the same purpose (either listen or transmit).
%
\end{enumerate}

\subsection{Paper Organization}
The rest of the paper is organized as follows.
Section~\ref{sec:sysModel} describes the general memoryless HD multi-relay network.
Section~\ref{sec:mainRes} first summarizes some known results for submodular functions and LPs, then proves the main result of the paper, and finally designs a polynomial-time algorithm to find the (approximately) optimal simple relay schedule.
Section~\ref{sec:Gaussian} applies the main result to Gaussian noise networks with multi-antenna nodes. In particular, we first show that NNC achieves the cut-set outer bound to within a constant gap that only depends on the total number of antennas, then we prove that the number of active states only depends on the number of relays (and not on the number of antennas) and we finally show that switching independently the antennas at each relay achieves higher rates than using all of them for the same purpose (either listen or transmit).
Section \ref{sec:Concl} concludes the paper.
Some proofs may be found in Appendix.

{

\subsection{Notation}
In the rest of the paper we use the following notation convention.  With $[n_1:n_2]$ we indicate the set of integers from $n_1$ to $n_2 \geq n_1$. For an index set $\mathcal{A}$ we let $Y_{\mathcal{A}} = \{ Y_j : j\in \mathcal{A} \}$. For two sets $\mathcal{A}_1,\mathcal{A}_2$, $\mathcal{A}_1 \subseteq \mathcal{A}_2$ indicates that $\mathcal{A}_1$ is a subset of $\mathcal{A}_2$, $\mathcal{A}_1 \cup \mathcal{A}_2$ represents the union of $\mathcal{A}_1$ and $\mathcal{A}_2$, while $\mathcal{A}_1 \cap \mathcal{A}_2$ represents the intersection of $\mathcal{A}_1$ and $\mathcal{A}_2$. With $\emptyset$ we denote the empty set and $\vert \mathcal{A} \vert $ indicates 
the cardinality of the set $\mathcal{A}$.
Lower and upper case letters indicate scalars, boldface lower case letters denote vectors and boldface upper case letters indicate matrices (with the exception of $Y^{j}$, which denotes a vector of length $j$ with components $(Y_1,\ldots,Y_j)$). $\mathbf{0}_j$ denotes the all-zero column vector of length $j$,
while $\mathbf{0}_{i \times j}$ is the all-zero matrix of dimension $i \times j$.
$\mathbf{1}_j$ is a column vector of length $j$ of all ones and $\mathbf{I}_j$ is the identity matrix of dimension $j$. $|\mathbf{A}|$ is the determinant of the matrix $\mathbf{A}$ and $\Tr \left[ \mathbf{A}\right ]$ is the trace of the matrix $\mathbf{A}$.
For a vector $\mathbf{a}$ we let ${\rm diag}[\mathbf{a}]$ be a diagonal matrix with the entries of $\mathbf{a}$ on the main diagonal, i.e., $\big[{\rm diag}[\mathbf{a}]\big]_{ij} = a_i \delta[i-j]$, where $\delta[n]$ 
is the Kronecker delta function. To indicate the block matrix $\mathbf{A}=\begin{bmatrix} \mathbf{A}_{1,1} & \mathbf{A}_{1,2} \\ \mathbf{A}_{2,1} & \mathbf{A}_{2,2} \end{bmatrix}$, we use the Matlab-inspired notation $\mathbf{A}=\left [\mathbf{A}_{1,1},\mathbf{A}_{1,2};\mathbf{A}_{2,1},\mathbf{A}_{2,2} \right]$; for the same block matrix $\mathbf{A}$, the notation $\mathbf{A}_{\mathcal{R},\mathcal{C}}$
indicates a submatrix of $\mathbf{A}$ where only the blocks in the rows indexed by the set $\mathcal{R}$
and the blocks in the columns indexed by the set $\mathcal{C}$ are retained.
$|a|$ is the absolute value of $a$ and $\| \mathbf{a} \|$ is the norm of the vector $\mathbf{a}$; $a^*$ is the complex conjugate of $a$, $\mathbf{a}^{T}$ is the transpose of the vector $\mathbf{a}$ and $\mathbf{a}^{\dagger}$ is the Hermitian transpose of the vector $\mathbf{a}$.
$X \sim \mathcal{N} \left ( \mu, \sigma^2 \right )$ indicates that $X$ is a proper-complex Gaussian random variable with mean $\mu$ and variance $\sigma^2$.
$\mathbb{E}\left [ \cdot \right ]$ indicates the expected value;
$[x]^+ := \max\{0,x\}$ for $x\in\mathbb{R}$ and $\log^+(a)= \max \{0,\log(a)\}$.

}

\section{System Model}
\label{sec:sysModel}

A memoryless relay network has one source (node 0), one destination (node $N+1$), and $N$ relays indexed from $1$ to $N$. It consists of $ N+1$ input alphabets $\left (\mathcal{X}_1,\cdots,\mathcal{X}_{N},\mathcal{X}_{N+1} \right )$ (here $\mathcal{X}_i$ is the input alphabet of node~$i$ except for the source / node~0 where, for notation convenience, we use $\mathcal{X}_{N+1}$ rather than $\mathcal{X}_{0}$), $N+1$ output alphabets $\left (\mathcal{Y}_{1},\cdots,\mathcal{Y}_{N},\mathcal{Y}_{N+1} \right )$ (here $\mathcal{Y}_i$ is the output alphabet of node~$i$), and a transition probability $\mathbb{P}_{Y_{[1:N+1]}|X_{[1:N+1]}}$.
The source has a message $W$ uniformly distributed on $[1:2^{n R}]$ for the destination, where $n$ denotes the codeword length and $R$ the transmission rate in bits per channel use (logarithms are in base $2$). 
At time $i$, $i \in [1:n]$, the source maps its message $W$ into a channel input symbol $X_{N+1,i}\left ( W \right )$, and the $k$-th relay, $k\in[1: N]$, maps its past channel observations into a channel input symbol $X_{k,i}\left ( Y_{k}^{i-1} \right )$. The channel is assumed to be memoryless, that is, the following Markov chain holds for all $i\in[1:n]$ 
\begin{align*}
    (W,Y_{[1: N+1]}^{i-1},X_{[1: N+1]}^{i-1}) 
\to X_{[1: N+1],i} 
\to Y_{[1: N+1],i}.
\end{align*}
At time $n$, the destination outputs an estimate of the message based on all its channel observations as $\widehat{W} \left (Y_{N+1}^n \right )$.
A rate $R$ is said to be $\epsilon$-achievable if there exists a sequence of codes indexed by the block length $n$ such that $\mathbb{P}[ \widehat{W} \neq W ] \leq \epsilon$ for some $\epsilon\in[0,1]$. The capacity is the largest non-negative rate  that is $\epsilon$-achievable for any $\epsilon>0$.

In this general memoryless framework, each relay can listen and transmit at the same time, i.e., it is a FD node.
HD channels are a special case of the memoryless FD framework in the following sense~\cite{kramer-allerton}.
With a slight abuse of notation compared to the previous paragraph, we let the channel input of the $k$-th relay, $k\in[1: N]$, be the pair $(X_k,S_k)$, where $X_k\in \mathcal{X}_k$ as before and $S_k \!\in\! [0:1]$ is the {\em state} random variable that indicates whether the $k$-th relay is in receive-mode ($S_k=0$) or in transmit-mode ($S_k=1$).
In the HD case the transition probability is specified as $\mathbb{P}_{Y_{[1:N+1]}|X_{[1:N+1]},S_{[1:N]}}$. In particular, when the $k$-th relay, $k\in [1:N]$, is listening ($S_k=0$) the outputs are independent of $X_k$, while when the $k$-th relay is transmitting ($S_k=1$) its output $Y_k$ is independent of all other random variables.

The capacity $\mathsf{C}$ of the HD multi-relay network is not known in general, but can be upper bounded by the cut-set bound 
\begin{align}
\mathsf{C} & \leq 
{ \max_{\mathbb{P}_{X_{[1:N+1]},S_{[1:N]}}}} \min_{\mathcal{A} \subseteq [1:N]} I_{\mathcal{A}}^{(\text{rand})},
\label{eq:capcutsetup}
\end{align}
where
\begin{align} 
I_{\mathcal{A}}^{(\text{rand})} &:=
I \left( X_{N + 1}, X_{\mathcal{A}^c},{ S_{\mathcal{A}^c}};
Y_{N + 1},Y_{\mathcal{A}} |X_{\mathcal{A}}, 
{ S_{\mathcal{A}}}
\right)
\label{eq:Iranddef}
\\&
\leq H({ S_{\mathcal{A}^c}}) + I_{\mathcal{A}}^{(\text{fix})}, 
\label{eq:Iranddefup}
\end{align}
for
\begin{align} 
I_{\mathcal{A}}^{(\text{fix})} &:=
I \left( X_{N + 1}, X_{\mathcal{A}^c};
Y_{N + 1},Y_{\mathcal{A}} |X_{\mathcal{A}},
{ S_{[1:N]}}
\right)
\label{eq:IAfixed}
\\&= \sum_{s  \in  [0:1]^N} \lambda_s \ f_s(\mathcal{A}),
\label{eq:IAfixedsumstate}
\end{align}
where 
\begin{align} 
\lambda_{s} &:= \mathbb{P}[S_{[1: N]}=s] \in[0,1] : \sum_{s\in [0:1]^N} \lambda_{s}=1, 
\label{eq:lambdas}
\\
f_s(\mathcal{A}) &:=
I \left( X_{N + 1}, X_{\mathcal{A}^c};
Y_{N + 1},Y_{\mathcal{A}} |X_{\mathcal{A}},  
{ S_{[1:N]} =s }
\right), \quad s\in [0:1]^N.
\label{eq:fs}
\end{align}
In the following, we use interchangeably the notation  $s  \in  [0:1]^N$ to index all possible binary vectors of length $N$, as well as, $s\in [0:2^N-1]$ to indicate the decimal representation of a binary vector of length $N$.
$I_{\mathcal{A}}^{(\text{rand})}$ in~\eqref{eq:Iranddef} is the mutual information across the network cut $\mathcal{A}\subseteq[1:N]$ when a {\it random schedule} is employed, i.e., information is conveyed from the relays to the destination by switching between listen and transmit modes of operation at random times~\cite{kramer-allerton} (see the term  $H({ S_{\mathcal{A}^c}})\leq \left |\mathcal{A}^c \right| \leq N$ in~\eqref{eq:Iranddefup}).
$I_{\mathcal{A}}^{(\text{fix})}$ in~\eqref{eq:IAfixed} is the mutual information with a {\it fixed schedule}, i.e., the time instants at which a relay transitions between listen and transmit modes of operation are fixed and known to all nodes in the network~\cite{kramer-allerton} (see the term $S_{[1:N]}$ in the conditioning in~\eqref{eq:IAfixed}).
Note that fixed schedules are optimal to within $N$ bits.


\section{Simple schedules for a class of HD multi-relay networks}
\label{sec:mainRes}

\begin{subequations}
We next consider networks for which the following holds:
there exists a product input distribution
\begin{align}
\mathbb{P}_{X_{[1:N+1]}|S_{[1:N]}} 
= \prod_{i\in[1:N+1]} \mathbb{P}_{X_i|S_{[1:N]}}
\label{eq:indipinputs}
\end{align}
for which we can evaluate the set function $I_{\mathcal{A}}^{(\text{fix})}$ in~\eqref{eq:IAfixed}
for all $\mathcal{A}\subseteq[1:N]$ and bound the capacity as
\begin{align}
\mathsf{C}^\prime-\mathsf{G}_1 &\leq  \mathsf{C}  \leq \mathsf{C}^\prime+\mathsf{G}_2, 
: \ \mathsf{C}^\prime := \max_{\mathbb{P}_{S_{[1:N]}} }  
\min_{\mathcal{A} \subseteq [1:N]}
I_{\mathcal{A}}^{(\text{fix})},
\label{eq:capinidinputs}
\end{align}
where $\mathsf{G}_1$ and $\mathsf{G}_2$ are non-negative constants that may depend on $N$ but not on the channel transition probability.
In other words, we concentrate on networks for which using independent inputs and a fixed relay schedule in the cut-set bound provides both an upper (to within $\mathsf{G}_2$ bits) and a lower (to within $\mathsf{G}_1$ bits) bounds on the capacity.

The main result of the paper is:
%
\begin{thm}\label{thm:main}
If in addition to the assumptions in~\eqref{eq:allassumptions} it also holds that
\begin{enumerate}
\item \label{thm:main:indpnoises}
the ``noises are independent,'' that is
\begin{align}
&\mathbb{P}_{Y_{[1:N+1]}|X_{[1:N+1]},S_{[1:N]}} = 
\prod_{i\in[1:N+1]} \mathbb{P}_{Y_{i}|X_{[1:N+1]},S_{[1:N]}},
\label{eq:indipnoises}
\end{align}
\item \label{thm:main:nolambdas}
and that the functions in~\eqref{eq:fs} are not a function of $\{\lambda_{s}, s\in[0:1]^N\}$, i.e., they can depend on the state $s$ but not on the $\{\lambda_{s}, s\in[0:1]^N\}$,
\end{enumerate}
then simple relay policies are optimal in~\eqref{eq:capinidinputs}, i.e., the optimal probability mass function $\mathbb{P}_{S_{[1:N]}}$ has at most $N+1$ non-zero entries / active states.
\end{thm}
\label{eq:allassumptions}
\end{subequations}

We first give some general definitions and 
summarize some
properties of submodular functions and LPs in Section~\ref{subsec:subLP},
we then prove Theorem~\ref{thm:main} in Sections~\ref{subsec:Thproof}-\ref{sec:proofStep3}, by also illustrating the different steps of the proof for the case $N=2$.
Finally, in Section~\ref{subsec:complexity} we discuss the computational complexity of finding (approximately) optimal simple schedules.

\subsection{Submodular Functions, LPs and Saddle-point Property} 
\label{subsec:subLP}

The following are standard results in submodular function optimization~\cite{BachSubm} and LPs~\cite{ChvatalLP}.

\begin{defin}[Submodular function, Lov\'{a}sz extension and greedy solution for submodular polyhedra]
\label{def:subm}
A set-function $f: 2^{N} \rightarrow \mathbb{R}$ is submodular if and only if, for all subsets $\mathcal{A}_1,\mathcal{A}_2 \subseteq [1:N]$, we have $f\left( \mathcal{A}_1\right)+f\left( \mathcal{A}_2\right) \geq f \left( \mathcal{A}_1 \cup \mathcal{A}_2 \right) + f \left( \mathcal{A}_1 \cap \mathcal{A}_2 \right)$
\footnote{A set-function $f$ is supermodular if and only if $-f$ is submodular, and
it is modular if it is both submodular and supermodular.}. 

Submodular functions are closed under non-negative linear combinations.

For a submodular function  $f$ such that $f(\emptyset) = 0$, the Lov\'{a}sz extension is the function $\widehat{f}:\mathbb{R}^N \to \mathbb{R}$  defined as
\begin{align}
\widehat{f} \left( \mathbf{w}\right) : = \max_{\mathbf{x} \in P(f)} \mathbf{w}^T \mathbf{x}, \quad  \forall\mathbf{w}  \in  \mathbb{R}^N,
\label{eq:lovext}
\end{align}
where $P(f)$ is the submodular polyhedron defined as 
\begin{align}
P(f):= \left\{ \mathbf{x} \in  \mathbb{R}^N : \sum_{i \in \mathcal{A}} x_i  \leq  f(\mathcal{A}), \ \forall \mathcal{A} \subseteq [1:N] \right\}.
\end{align}
The optimal $\mathbf{x}$ in~\eqref{eq:lovext} can be found by the greedy algorithm for submodular polyhedra and has components 
\begin{align}
x_{\pi_i}= f\left( \{\pi_1, \ldots , \pi_i\}\right) -f\left( \{\pi_1, \ldots , \pi_{i-1}\}\right), \forall i \in [1:N],
\end{align}
where $\pi$ is a permutation of $[1:N]$ such that the weights $\mathbf{w}$ are ordered as $w_{\pi_1} \geq w_{\pi_2} \geq \ldots \geq w_{\pi_N}$, and where by definition $\{\pi_{0}\} = \emptyset$.  

The Lov\'{a}sz extension is a piecewise linear convex function.
\end{defin}

\begin{prop}[Minimum of submodular functions]
\label{prop:minsub}
Let $f$ be a submodular function 
and $\widehat{f}$ its Lov\'{a}sz extension.
The minimum of the submodular function satisfies
\begin{align*}
\min_{\mathcal{A} \subseteq [1:N]} f \left( \mathcal{A} \right) = \min_{\mathbf{w} \in [ 0:1]^N } \widehat{f} \left( \mathbf{w} \right)= \min_{\mathbf{w} \in [ 0,1]^N } \widehat{f} \left( \mathbf{w} \right),
\end{align*}
i.e., $\widehat{f} \left( \mathbf{w} \right)$ attains its minimum at a vertex of the cube $[0,1]^N$.
\end{prop}

\begin{defin}[Basic feasible solution]
\label{def:BFS}
Consider the LP
\begin{align*}
\begin{array}{ll}
{\rm{maximize}} & \mathbf{c}^T \mathbf{x} 
\\ {\rm{subject \ to}} & \mathbf{A} \mathbf{x} \leq \mathbf{b} \quad \mathbf{x} \geq 0,
\end{array}
\end{align*}
where $\mathbf{x}\in  \mathbb{R}^n$ is the vector of unknowns, 
$\mathbf{b}\in  \mathbb{R}^m$ and $\mathbf{c}\in  \mathbb{R}^n$ are vectors of known coefficients,
and $\mathbf{A}\in  \mathbb{R}^{m \times n}$ is a known matrix of coefficients.
If $m<n$, a solution for the LP with at most $m$ non-zero values is called a basic feasible solution.
\end{defin}

\begin{prop}[Optimality of basic feasible solutions]
\label{th:optsolextr}
If a LP is feasible, then an optimal solution is at a vertex of the (non-empty and convex) feasible set $S= \left \{ \mathbf{x} \in \mathbb{R}^{n}: \mathbf{A}\mathbf{x} \leq \mathbf{b}, \mathbf{x} \geq 0\right \}$.
Moreover, if there is an optimal solution, then an optimal basic feasible solution exists as well.
\end{prop}

\begin{prop}[Saddle-point property]
\label{prop:minmaxeq}
Let $\phi(x,y)$ be a function of two vector variables $x \in \mathcal{X}$ and $y \in \mathcal{Y}$. By  the minimax inequality we have
\begin{align*}
\max_{y \in \mathcal{Y}}\min_{x \in \mathcal{X}} \phi \left( x,y\right) \leq \min_{x \in \mathcal{X}} \max_{y \in \mathcal{Y}} \phi \left( x,y\right)
\end{align*}
and equality holds 
if the following three conditions hold:
(i) $\mathcal{X}$ and $\mathcal{Y}$ are both convex and one of them is compact, 
(ii) $\phi \left( x,y\right)$ is convex in $x$ and concave in $y$, and
(iii) $\phi \left( x,y\right)$ is continuous.
\end{prop}

\subsection{Overview of the Proof of Theorem \ref{thm:main}}
\label{subsec:Thproof}
The objective is to show that simple relay policies are optimal in~\eqref{eq:capinidinputs}.
The proof consists of the following steps:
\begin{enumerate}
\item
We first show that the function $I_{\mathcal{A}}^{(\text{fix})}$ defined in~\eqref{eq:IAfixed} is submodular under the assumptions in \eqref{eq:allassumptions}. 
\item
By using Proposition~\ref{prop:minsub}, we show that the problem in~\eqref{eq:capinidinputs} can be recast into an equivalent max-min problem.
\item
With Proposition~\ref{prop:minmaxeq} we show that the  max-min~problem is equivalent to solve a min-max~problem.
The min-max~problem is then shown to be equivalent to solve $N!$ max-min problems, for each of which we obtain an optimal basic feasible solution by Proposition~\ref{th:optsolextr} with the claimed maximum number of non-zero entries.
\end{enumerate}
We now give the details for each step in a separate subsection.

\subsection{Proof Step 1}
%
We show that $I_{\mathcal{A}}^{(\text{fix})}$ in~\eqref{eq:IAfixed} is submodular.
The result in~\cite[Theorem 1]{ParvaEtk} showed that $f_s(\mathcal{A})$ in~\eqref{eq:fs} is submodular for each relay state $s\in  [0:1]^N$ under the assumption of independent inputs and independent noises (the same work provides an example of a diamond network with correlated  inputs for which the cut-set bound is neither submodular nor supermodular). Since submodular functions are closed under non-negative linear combinations (see Definition~\ref{def:subm}), this implies that $I_{\mathcal{A}}^{(\text{fix})}  = \sum_{s  \in  [0:1]^N} \lambda_s \ f_s(\mathcal{A})$ is submodular under the assumptions of Theorem~\ref{thm:main}. For completeness, we provide the proof of this result in Appendix \ref{app:submodularityProof}, where we use Definition~\ref{def:subm} as opposed to 
the ``diminishing marginal returns'' property of a submodular function used in \cite{ParvaEtk}.

\subsubsection*{Example for $N=2$} 
In this setting we have $2^2=4$ possible cuts, each of which is a linear combination of $2^2=4$ possible listen/transmission configuration states. In particular, from \eqref{eq:IAfixedsumstate} we have
\begin{align*}
\begin{array}{ll}
\mathcal{A}= \emptyset, & I_{\emptyset}^{(\text{fix})}:= \lambda_0 f_0 \left (\emptyset \right ) + \lambda_1 f_1 \left (\emptyset \right )+\lambda_2 f_2 \left (\emptyset \right )+\lambda_3 f_3 \left (\emptyset \right ),
\\
\mathcal{A}= \left \{ 1 \right \}, & I_{\left \{ 1 \right \}}^{(\text{fix})}:= \lambda_0 f_0 \left (\left \{ 1 \right \} \right ) + \lambda_1 f_1 \left (\left \{ 1 \right \} \right )+\lambda_2 f_2 \left (\left \{ 1 \right \} \right )+\lambda_3 f_3 \left (\left \{ 1 \right \} \right ),
\\
\mathcal{A}= \left \{ 2 \right \}, & I_{\left \{ 2 \right \}}^{(\text{fix})}:= \lambda_0 f_0 \left (\left \{ 2 \right \} \right ) + \lambda_1 f_1 \left (\left \{ 2 \right \} \right )+\lambda_2 f_2 \left (\left \{ 2 \right \} \right )+\lambda_3 f_3 \left (\left \{ 2 \right \} \right ),
\\
\mathcal{A}= \left \{ 1,2 \right \}, & I_{\left \{ 1,2 \right \}}^{(\text{fix})}:= \lambda_0 f_0 \left (\left \{ 1,2 \right \} \right ) + \lambda_1 f_1 \left (\left \{ 1,2 \right \} \right )+\lambda_2 f_2 \left (\left \{ 1,2 \right \} \right )+\lambda_3 f_3 \left (\left \{ 1,2 \right \} \right ),
\end{array}
\end{align*}
where, $\forall s \in [0:3]$, we have that the functions in~\eqref{eq:fs} are given by
\begin{align*}
f_s \left (\emptyset \right ) &: = I \left(X_{3},X_{2},X_{1}; Y_{3}| S_{[1:2]}=s\right),
\\
f_s \left (\left \{ 1\right \}\right ) &: = I \left( X_{3},X_{2}; Y_{3},Y_{1}|X_{1}, S_{[1:2]}=s \right),
\\
f_s \left (\left \{ 2\right \}\right ) &: = I \left( X_{3},X_{1}; Y_{3},Y_{2}|X_{2}, S_{[1:2]}=s \right),
\\
f_s \left (\left \{ 1,2\right \}\right ) &: = I \left( X_{3}; Y_{3},Y_{2},Y_{1}|X_{2},X_{1}, S_{[1:2]}=s \right),
\end{align*}
and are submodular under the assumptions in~\eqref{eq:allassumptions}.

\subsection{Proof Step 2}
Given that $I_{\mathcal{A}}^{(\text{fix})}$ in \eqref{eq:IAfixed} is submodular, we would like to use Proposition~\ref{prop:minsub} to replace the minimization over the subsets of $[1:N]$ in~\eqref{eq:capinidinputs} with a minimization over the cube $[0:1]^N$. Since $I_{\emptyset}^{(\text{fix})} = I \left( X_{[1:N+1]}; Y_{N+1}| S_{[1:N]} \right)\geq 0$ in general, we define a new submodular function 
\begin{align}
g\left( \mathcal{A} \right) := I_{\mathcal{A}}^{(\text{fix})} - I_{\emptyset}^{(\text{fix})}
\end{align}
and proceed as follows
\begin{align}
\min_{\mathcal{A} \subseteq [1:N]} I_{\mathcal{A}}^{(\text{fix})}
&= I_{\emptyset}^{(\text{fix})} + \min_{\mathcal{A} \subseteq [1:N]} g\left( \mathcal{A} \right) \nonumber
\\&
= I_{\emptyset}^{(\text{fix})} + \min_{\mathbf{w}\in [0,1]^N} 
\begin{bmatrix} w_{\pi_1} & w_{\pi_2} & \ldots & w_{\pi_N} \\ \end{bmatrix}
\begin{bmatrix} 
g\left( \{\pi_1\} \right) - g\left( \emptyset \right) \\ 
\vdots \\ 
g\left( \{\pi_1, \ldots , \pi_N\} \right) -g\left( \{\pi_1, \ldots , \pi_{N-1}\} \right) \\ \end{bmatrix} \nonumber
\\&= I_{\emptyset}^{(\text{fix})} +  \min_{\mathbf{w}\in [0,1]^N}
\begin{bmatrix} w_{\pi_1} & w_{\pi_2} & \ldots & w_{\pi_N} \\ \end{bmatrix}
\begin{bmatrix} 
I_{\{\pi_1\}}^{(\text{fix})}- I_{\emptyset}^{(\text{fix})} \\ 
\vdots \\ 
I_{\{\pi_1, \ldots , \pi_N\}}^{(\text{fix})} -I_{\{\pi_1, \ldots , \pi_{N-1}\}}^{(\text{fix})} \\ \end{bmatrix} \nonumber
\\&=   \min_{\mathbf{w}\in [0,1]^N}
\begin{bmatrix} 1 & w_{\pi_1} & w_{\pi_2} & \ldots & w_{\pi_N} \\ \end{bmatrix}
\begin{bmatrix} 
I_{\emptyset}^{(\text{fix})} \\
I_{\{\pi_1\}}^{(\text{fix})}  - I_{\emptyset}^{(\text{fix})} \\ 
\vdots \\ 
I_{\{\pi_1, \ldots , \pi_N\}}^{(\text{fix})} -I_{\{\pi_1, \ldots , \pi_{N-1}\}}^{(\text{fix})} \\ \end{bmatrix}  \nonumber
\\&=: \min_{\mathbf{w}\in [0,1]^N} \left\{ [ 1, \mathbf{w}^T ] \ \mathbf{H}_{\pi,f} \right\},
\label{eq:optle}
\end{align}
which implies that the problem in~\eqref{eq:capinidinputs} is equivalent to
\begin{align}
\mathsf{C}^\prime = 
\max_{{\bf \lambda}_{\rm{vect}}} \min_{\mathbf{w}\in [0,1]^N} \Big\{ [ 1, \mathbf{w}^T ] \ \mathbf{H}_{\pi,f} {\bf \lambda}_{\rm{vect}} \Big\},
\label{eq:capacequivalent}
\end{align}
where ${\bf \lambda}_{\rm{vect}}$ is the probability mass function of $S_{[1:N]}$ in~\eqref{eq:lambdas}, 
$\mathbf{H}_{\pi,f}$ is defined as
\begin{align}
\mathbf{H}_{\pi,f}
&:=
\mathbf{P}_{\pi} \
\underbrace{
\begin{bmatrix}
 1 & 0 & 0 & \ldots & 0\\
-1 & 1 & 0 & \ldots & 0\\
 0 &-1 & 1 & \ldots & 0\\
\vdots \\
 0 & 0& \ldots &-1 & 1\\
\end{bmatrix} 
}_{(N+1) \times (N+1)} \mathbf{F}_{\pi} \in \mathbb{R}^{(N+1) \times 2^N},
\label{eq:defHpif}
\end{align}
where $\mathbf{P}_{\pi}  \in  \mathbb{R}^{(N + 1) \times (N + 1)}$ is the permutation matrix that maps
$[ 1 , w_{1} , \ldots , w_{N}]$ into $[1 , w_{\pi_1} , \ldots , w_{\pi_N}]$,
and $\mathbf{F}_{\pi}$ is defined as
\begin{align}
\mathbf{F}_{\pi}&:=
\begin{bmatrix}
f_0(\emptyset)         & \ldots & f_{2^N-1}(\emptyset)         \\
f_0(\{\pi_1\})             & \ldots & f_{2^N-1}(\{\pi_1\})             \\
f_0(\{\pi_1,\pi_2\})       & \ldots & f_{2^N-1}(\{\pi_1,\pi_2\})       \\
\ldots                       \\
f_0(\{\pi_1, \ldots, \pi_N\}) & \ldots & f_{2^N-1}(\{\pi_1, \ldots, \pi_N\}) \\
\end{bmatrix}  \in \mathbb{R}^{(N+1) \times 2^N},
\end{align}
with $f_s \left( \mathcal{A}\right)$ being defined in~\eqref{eq:fs}.
We thus expressed our original optimization problem in~\eqref{eq:capinidinputs} as the max-min~problem in~\eqref{eq:capacequivalent}.

\subsubsection*{Example for $N=2$} 
\begin{figure}
\centering
\includegraphics[width=0.7\textwidth]{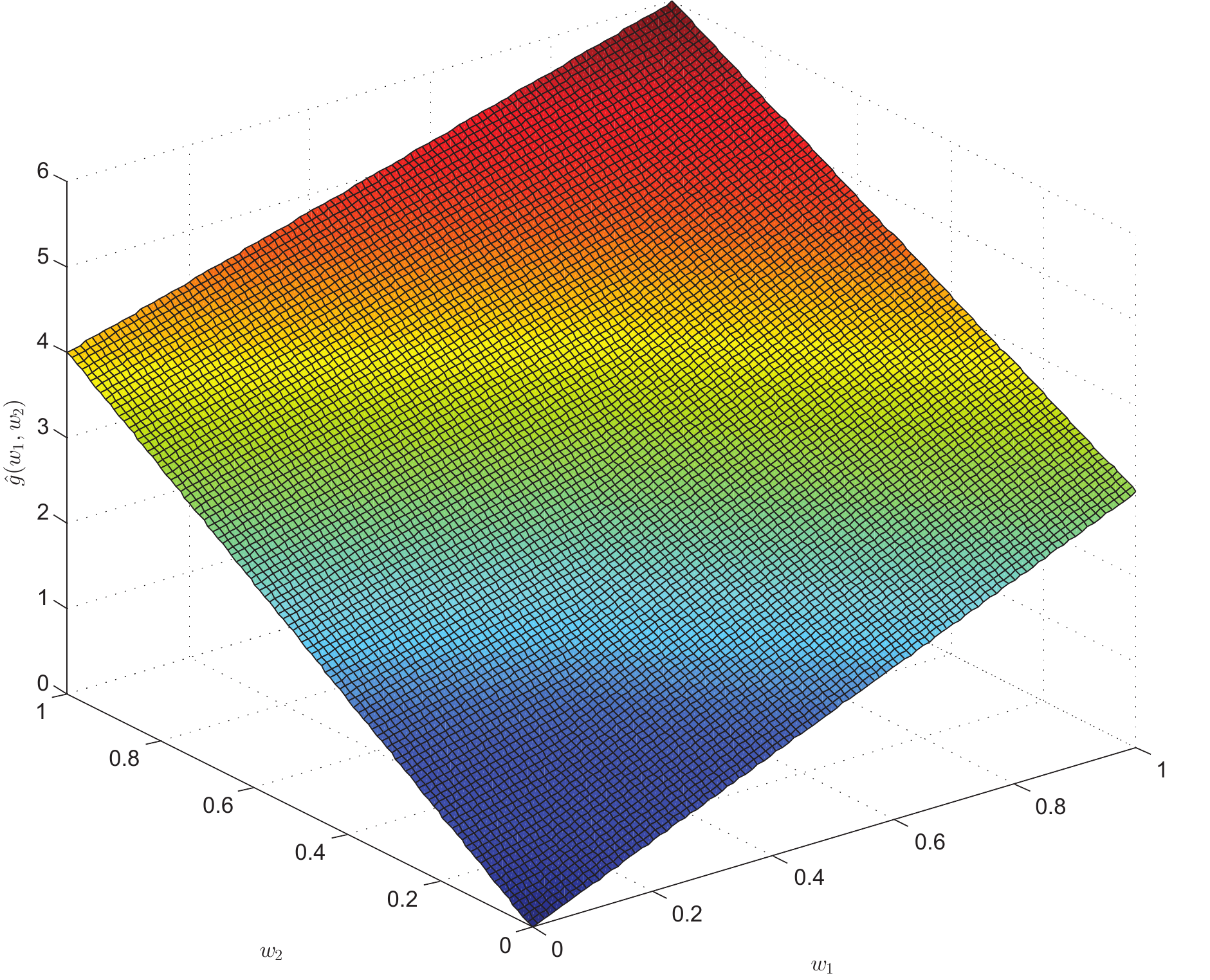}
\caption{Lov\'{a}sz extension $\widehat{g}(w_1,w_2) $ in \eqref{eq:leN2}, with $g \left( \left \{1 \right \}\right)=3$, $g \left( \left \{2 \right \}\right)=4$ and $g \left( \left \{1,2 \right \}\right)=6$. }
\label{fig:LE}
\end{figure}

With $N=2$, we have $g \left( \mathcal{A} \right) = I_{\mathcal{A}}^{(\text{fix})} - I_{\emptyset}^{(\text{fix})},  \mathcal{A}\subseteq[1:2]$ and the Lov\'{a}sz extension (see Definition \ref{def:subm}) is
\begin{align}
\label{eq:leN2}
\widehat{g}(w_1,w_2) =
 \left \{ \begin{array}{cc}  
  w_1 g \left( \left \{1 \right \}\right) 
+ w_2 \left[  g \left( \left \{1,2 \right \}\right) - g \left( \left \{1 \right \}\right)\right] & \text{if} \ w_1 \geq w_2 
\\ 
 w_2 g \left( \left \{2 \right \}\right )
+ w_1 \left[  g \left( \left \{1,2 \right \}\right) - g \left( \left \{2\right \}\right)\right] & \text{if} \ w_2 \geq w_1 \end{array} \right. .
\end{align}
A visual representation of the Lov\'{a}sz extension $\widehat{g}(w_1,w_2)$ in \eqref{eq:leN2} on  $[0,1]^2$ is given in Fig. \ref{fig:LE}, where we considered $g \left( \left \{1 \right \}\right)=3$, $g \left( \left \{2 \right \}\right)=4$ and $g \left( \left \{1,2 \right \}\right)=6$ (recall $g(\emptyset)=0$).

Let
\begin{align}
i_{{\rm{M}}} &:= \arg \max \left \{ w_1, w_2\right \} \quad {\text{and}} \quad i_{{\rm{m}}} := \arg \min \left \{ w_1, w_2\right \}.
\label{eq:MaxMinInde}
\end{align}
The optimization problem in \eqref{eq:optle} for $N=2$ can be written as 
\begin{align}
& \min_{0 \leq w_{i_{{\rm{m}}}} \leq w_{i_{{\rm{M}}}} \leq 1 }
\left \{
\begin{bmatrix} 1 & w_{i_{{\rm{M}}}} & w_{i_{{\rm{m}}}}\end{bmatrix}
\begin{bmatrix}
 1 & 0 & 0 \\
-1 & 1 & 0\\
 0 &-1 & 1\\
\end{bmatrix}
\mathbf{F}_{\pi} \right \} \notag
\\&=\min_{0 \leq w_{i_{{\rm{m}}}} \leq w_{i_{{\rm{M}}}} \leq 1 }
\left \{ \begin{bmatrix} 1-w_{i_{{\rm{M}}}} & w_{i_{{\rm{M}}}}-w_{i_{{\rm{m}}}} & w_{i_{{\rm{m}}}} \end{bmatrix}
\mathbf{F}_{\pi} \right \},
\label{eq:optimN2}
\end{align}
with
\begin{align}
\mathbf{F}_{\pi}
&= \begin{bmatrix}
f_0(\emptyset) & f_1(\emptyset) & f_{2}(\emptyset) & f_{3}(\emptyset) \\
f_0(\{i_{{\rm{M}}}\})         & f_1(\{i_{{\rm{M}}}\})         & f_{2}(\{i_{{\rm{M}}}\})         & f_{3}(\{i_{{\rm{M}}}\})         \\
f_0(\{1,2\})       & f_1(\{1,2\})       & f_{2}(\{1,2\})       & f_{3}(\{1,2\})       \\
\end{bmatrix},
\end{align}
and finally 
the optimization problem in \eqref{eq:capacequivalent} is
\begin{align}
&\mathsf{C}^\prime = 
\max_{{\bf \lambda}_{\rm{vect}}} \min_{0 \leq w_{i_{{\rm{m}}}} \leq w_{i_{{\rm{M}}}} \leq 1 } 
\left \{
\begin{bmatrix} 1-w_{i_{{\rm{M}}}} & w_{i_{{\rm{M}}}}-w_{i_{{\rm{m}}}} & w_{i_{{\rm{m}}}} \end{bmatrix}
\mathbf{F}_{\pi}
\begin{bmatrix} \lambda_0 \\ \lambda_1 \\ \lambda_2 \\ \lambda_3  \\ \end{bmatrix} \right \}.
\label{eq:capacequivalentN2}
\end{align}

\subsection{Proof Step 3}
\label{sec:proofStep3}
In order to solve~\eqref{eq:capacequivalent} we would like to reverse the order of $\min$ and $\max$.
We note that the function $\phi \left( {\bf \lambda}_{\rm{vect}}, \mathbf{w}\right) := [ 1, \mathbf{w}^T ] \ \mathbf{H}_{\pi,f} {\bf \lambda}_{\rm{vect}}$ satisfies the properties in Proposition~\ref{prop:minmaxeq} (it is continuous; it is convex in $\mathbf{w}$ by the convexity of the Lov\'{a}sz extension and linear (under the assumption in item~\ref{thm:main:nolambdas} in Theorem~\ref{thm:main}), thus concave, in ${\bf \lambda}_{\rm{vect}}$; the optimization domain in both variables is compact).
Thus, we now focus on the problem
\begin{align}
\mathsf{C}^\prime=\min_{\mathbf{w}\in [0,1]^N} \max_{{\bf \lambda}_{\rm{vect}}}
\Big\{ [ 1, \mathbf{w}^T ] \ \mathbf{H}_{\pi,f} {\bf \lambda}_{\rm{vect}} \Big\},
\end{align}
which can be equivalently rewritten as
\begin{align}
\mathsf{C}^\prime
&=\min_{\pi \in \mathcal{P}_N} \min_{\mathbf{w}_{\pi}\in [0:1]^N} \max_{{\bf \lambda}_{\rm{vect}}}
\Big\{ [ 1, \mathbf{w}_{\pi}^T ] \ \mathbf{H}_{\pi,f} {\bf \lambda}_{\rm{vect}} \Big\}
\label{eq:P1}
\\&=\min_{\pi \in \mathcal{P}_N}  \max_{{\bf \lambda}_{\rm{vect}}}\min_{\mathbf{w}_{\pi}\in [0:1]^N}
\Big\{ [ 1, \mathbf{w}_{\pi}^T ] \ \mathbf{H}_{\pi,f} {\bf \lambda}_{\rm{vect}} \Big\},
\label{eq:P2}
\end{align}
where $\mathcal{P}_N$ is the set of all the $N!$ permutations of $[1:N]$.
In~\eqref{eq:P1}, for each permutation $\pi \in \mathcal{P}_N$, 
we first find the optimal ${\bf \lambda}_{\rm{vect}}$, 
and then find the optimal $\mathbf{w}_{\pi} : w_{\pi_1} \geq w_{\pi_2} \geq \ldots w_{\pi_N}$.
This is equivalent to~\eqref{eq:P2}, where again by Proposition~\ref{prop:minmaxeq}, 
for each permutation $\pi \in \mathcal{P}_N$, 
we first find the optimal  $\mathbf{w}_{\pi} : w_{\pi_1} \geq w_{\pi_2} \geq \ldots w_{\pi_N}$,
and then find the optimal ${\bf \lambda}_{\rm{vect}}$. 

Let now consider the inner optimization in~\eqref{eq:P2}, that is, the problem
\begin{align}
P_1: 
\max_{{\bf \lambda}_{\rm{vect}}} \min_{\mathbf{w}_{\pi}\in [0:1]^N}  \Big\{ [ 1, \mathbf{w}_{\pi}^T ] \ \mathbf{H}_{\pi,f} {\bf \lambda}_{\rm{vect}} \Big\}.
\label{eq:newOptPrDual}
\end{align}
From Proposition \ref{prop:minsub} we know that, for a given $\pi \in \mathcal{P}_N$, the optimal $\mathbf{w}_{\pi}$ is a vertex of the cube $[0:1]^N$. For a given ${\pi} \in \mathcal{P}_N$, there are $N+1$ vertices whose coordinates are ordered according to $\pi$. 
%
In~\eqref{eq:newOptPrDual}, for each of the $N+1$ feasible vertices of $\mathbf{w}_{\pi}$, it is easy to see that the product $[ 1, \mathbf{w}_{\pi}^T ] \ \mathbf{H}_{\pi,f}$ is equal to a row of the matrix $\mathbf{F}_{\pi}$. By considering all possible $N+1$ feasible vertices compatible with $\pi$ we obtain all the $N+1$ rows of the matrix $\mathbf{F}_{\pi}$.
Hence, 
$P_1$ is equivalent to
\begin{align}
\begin{array}{lll}
P_2:&{\rm{maximize}} & \tau 
\\ &{\rm{subject \ to}} &  \mathbf{1}_{(N+1)} \tau  \leq \mathbf{F}_{\pi} {\bf \lambda}_{\rm{vect}}
\\ &{\rm{and}} &  \mathbf{1}_{2^N}^T  {\bf \lambda}_{\rm{vect} } = 1, \ {\bf \lambda}_{\rm{vect} }\geq \mathbf{0}_{2^N}, \ \tau \geq 0.
\end{array}
\label{eq:polmnuytf}
\end{align}
The LP $P_2$ in~\eqref{eq:polmnuytf} has 
$n=2^N+1$ optimization variables ($2^N$ values for ${\bf \lambda}_{\rm{vect}}$ and one value for $\tau$),
$m=N+2$ constraints, and 
is feasible (consider for example 
the uniform distribution of ${\bf \lambda}_{\rm{vect}}$ and  $\tau=0$).
Therefore, by Proposition~\ref{th:optsolextr}, $P_2$ has an optimal basic feasible solution with at most $m=N+2$ non-zero values. Since $\tau >0$ (otherwise the channel capacity would be zero), it means that ${\bf \lambda}_{\rm{vect}}$ has at most $N+1$ non-zero entries.

Since for each $\pi \in \mathcal{P}_N$ the optimal ${\bf \lambda}_{\rm{vect}}$ in~\eqref{eq:P2} has at most $N+1$ non-zero values, then also for the optimal permutation the corresponding optimal ${\bf \lambda}_{\rm{vect}}$ has at most $N+1$ non-zero values. This shows that the (approximately) optimal schedule in the original problem in~\eqref{eq:capinidinputs} is simple. 

This concludes the proof of Theorem \ref{thm:main}.

\subsubsection*{Example for $N=2$} 
For $N=2$, we have $|\mathcal{P}_2|= 2!=2$ possible permutations.
From Proposition \ref{prop:minsub}, the optimal $\mathbf{w}$
is one of the vertices $(0,0),(0,1),(1,0),(1,1)$. Let now focus on the case $i_{{\rm{M}}}=1$ and $i_{{\rm{m}}}=2$ (a similar reasoning holds for $i_{{\rm{M}}}=2$ and $i_{{\rm{m}}}=1$ as well). 
Under this condition $P_1$ in \eqref{eq:newOptPrDual} is the problem in \eqref{eq:capacequivalentN2} with $i_{{\rm{M}}}=1$ and $i_{{\rm{m}}}=2$. The vertices compatible with this permutation are
$(w_1,w_2)\in\{(0,0),(1,0),(1,1) \}$, 
which result in $(1-w_1, w_1-w_2, w_2)\in\{(0,0,0),(0,1,0),(0,0,1) \}$. 
This implies that $P_2$ in \eqref{eq:polmnuytf} is
\begin{align}
\begin{array}{lll}
P_2: &{\rm{maximize}} & \tau \\
     &{\rm{subject \ to}} 
     & \tau  \leq f_0(\emptyset)\lambda_0 + f_1(\emptyset)\lambda_1 + f_{2}(\emptyset)\lambda_2 + f_{3}(\emptyset)\lambda_3, \\
    && \tau  \leq f_0(\{1\})\lambda_0 + f_1(\{1\})\lambda_1 + f_{2}(\{1\})\lambda_2 + f_{3}(\{1\})\lambda_3, \\
    && \tau  \leq f_0(\{1,2\})\lambda_0 + f_1(\{1,2\})\lambda_1 + f_{2}(\{1,2\})\lambda_2 + f_{3}(\{1,2\})\lambda_3, \\
    && \lambda_0 + \lambda_1 + \lambda_2 + \lambda_3 = 1, \ \lambda_i\geq 0 \ i\in[0:3], \ \tau\geq 0,
\end{array}
\label{eq:P2equiv N=2 w1>w2}
\end{align}
where each of the three inequality constraints correspond to a different row of $\mathbf{F}_{\pi}$
multiplied by ${{\bf \lambda}_{\rm{vect}}}=[\lambda_0,\lambda_1,\lambda_2,\lambda_3]^T$. 
Therefore, $P_2$ in~\eqref{eq:P2equiv N=2 w1>w2} has four constraints (three from the rows of $\mathbf{F}_{\pi}$ and one from ${{\bf \lambda}_{\rm{vect}}}$) and five unknowns (one value for $\tau$ and four entries of ${{\bf \lambda}_{\rm{vect}}}$). Thus, by Proposition~\ref{th:optsolextr}, $P_2$ has an optimal basic feasible solution with at most four non-zero values, of which one is $\tau$ and thus the other (at most) three belong to ${{\bf \lambda}_{\rm{vect}}}$. 

By~\cite[Appendix C]{ourITjournal}, we know that either $\lambda_0$ or $\lambda_3$ is zero, thus giving the desired (approximately) optimal simple schedule. 

{
\begin{rem} \rm
\label{rem:nolindip}
In order to apply the saddle-point property (see Proposition \ref{prop:minmaxeq}) and hence cast our optimization problem as a LP, the proof of Step 3 requires that the matrix $\mathbf{F}_{\pi}$ does not depend on ${\bf \lambda}_{\rm{vect}}$; this is the reason of our assumption in item~\ref{thm:main:nolambdas} in Theorem~\ref{thm:main}.
In our Gaussian noise example (see Section \ref{sec:Gaussian}), this excludes the possibility of power allocation across the relay states because power allocation makes the optimization problem non-linear in ${\bf \lambda}_{\rm{vect}}$. 
\end{rem}

\begin{rem} \rm
As stated in Theorem \ref{thm:main}, the assumptions in~\eqref{eq:allassumptions} provide a set of sufficient 
conditions for the existence of an (approximately) optimal simple schedule. Since those conditions are not necessary,  there might exist networks for which the assumptions in~\eqref{eq:allassumptions} are not satisfied, but for which the (approximately) optimal schedule is still simple. Determining necessary conditions for optimality of simple schedules is an interesting challenging open question.
\end{rem}

\begin{rem} \rm
For FD relays, it was showed in~\cite{ParvaEtk} that wireless erasure networks, Gaussian networks with single-antenna nodes and their linear deterministic high-SNR approximations are examples for which the cut-set bound (or an approximation to it) is submodular. Since submodular functions are closed under non-negative linear combinations (see Definition~\ref{def:subm}), this implies that the cut-set bound (or an approximation to it) is still submodular when evaluated for these same networks with HD relays. As a consequence, Theorem \ref{thm:main} holds for wireless erasure networks, Gaussian networks with single-antenna nodes and their linear deterministic high-SNR approximations with HD relays.
\end{rem}
}

\subsection{On the complexity of finding the  (approximately) optimal simple schedule}
\label{subsec:complexity}
Our proof method for Theorem \ref{thm:main} seems to suggest that finding the  (approximately) optimal schedule requires the solution of $N!$ different LPs.
Since $\log(N!) = O(N\log(N/e))$, the computational complexity of such an approach would be prohibitive for large $N$. 
Next we propose a polynomial-time algorithm in $N$ to determine the  (approximately) optimal simple schedule for any network regardless of its connectivity / topology.

The idea is to use an iterative method that alternates between a submodular function minimization over $\mathbf{w}$  and a LP maximization over ${\bf \lambda}_{\rm{vect}}$. The saddle-point property in Proposition \ref{prop:minmaxeq}, which holds with equality in our setting, ensures that the algorithm converges to the optimal solution. 
The pseudo-code of the proposed algorithm 
is given below.
The algorithm runs in polynomial-time since:
\begin{enumerate}
\item[a)] the unconstrained minimization of our submodular function can be solved in strongly polynomial-time in $N$;
in particular, the algorithm in \cite{Orlin09} runs in $O \left( N^5\kappa+N^6\right)$, with $\kappa$ being the time the algorithm needs to compute $f_s(\mathcal{A})$ in \eqref{eq:fs} for any subset $\mathcal{A} \subseteq [1:N]$ and for each state $s \in [0:1]^N$;
\item[b)] by strong duality, the dual of our LP maximization in~\eqref{eq:capacequivalent} with $N+2$ unknowns can be solved in polynomial-time in $N$; in particular, the ellipsoid method in \cite{GLS1984c} has complexity $O\left(N^4 \right )$. 
\end{enumerate}
%

\begin{algorithm}
  \caption{Find $\mathsf{C}^\prime$ in \eqref{eq:capacequivalent}}
  \KwIn{Matrix $\mathbf{H}_{\pi,f}$ defined in \eqref{eq:defHpif}, ${\rm{MyToll}}$}
  \KwOut{$\mathsf{C}^\prime$, $\mathbf{w}$ and ${\bf \lambda}_{\rm{vect}}$ in \eqref{eq:capacequivalent}}
${\rm{t}}=0$\;
${\bf \lambda}_{\rm{vect}}\left[t \right]= \frac{1}{2^N} {\mathbf{1}_{2^N}}$\;
$\mathbf{w}\left[t \right]= \mathbf{0}_N$\;
    \While{${\rm{err}}> {\rm{MyToll}}$}
    {
    ${\rm{t}} \leftarrow {\rm{t}} +1$\;
    $\left( \mathsf{C}^\prime_{\mathbf{w}},\mathbf{w}[t] \right ) \leftarrow$solve $\min_{\mathbf{w}} \Big\{ [ 1, \mathbf{w}^T ] \ \mathbf{H}_{\pi,f} {\bf \lambda}_{\rm{vect}}\left[t-1 \right] \Big\}$\;
$\left( \mathsf{C}^\prime_{{\bf \lambda}_{\rm{vect}}},{\bf \lambda}_{\rm{vect}}[t] \right ) \leftarrow$ solve $\max_{{\bf \lambda}_{\rm{vect}}} \Big\{ [ 1, \mathbf{w}^T[t] ] \ \mathbf{H}_{\pi,f} {\bf \lambda}_{\rm{vect}}\Big\}$\;
${\rm{err}} \leftarrow \left| \mathsf{C}^\prime_{\mathbf{w}} - \mathsf{C}^\prime_{{\bf \lambda}_{\rm{vect}}}\right|$\;
    }
\Return $\mathsf{C}^\prime_{\mathbf{w}},\mathbf{w}[t],{\bf \lambda}_{\rm{vect}}[t]$.
\label{algo:maxminSP}
\end{algorithm}

\section{Example: the Gaussian noise case with multi-antenna nodes}
\label{sec:Gaussian} 

In this section we show that Theorem~\ref{thm:main} applies to the practically relevant Gaussian noise network where the nodes are equipped with multiple antennas and where the $N$ relays operate in HD mode.
The complex-valued power-constrained Gaussian MIMO HD relay network has input/output relationship 
\begin{subequations}
\begin{align}
&\mathbf{y} =\mathbf{H}_{\rm eq} \mathbf{x} +\mathbf{z} \in \mathbb{C}^{(m_{\rm{tot}} +m_{N+1})\times 1},\\
&\mathbf{H}_{\rm eq}
:=
\begin{bmatrix}
 \mathbf{I}_{m_{\rm{tot}}}-\mathbf{S} & \mathbf{0}_{m_{\rm{tot}} \times m_{N+1}} \\
  \mathbf{0}_{m_{N+1} \times m_{\rm{tot}}}  & \mathbf{I}_{m_{N+1}} \\
 \end{bmatrix}
\ \mathbf{H} \
\begin{bmatrix}
 \mathbf{S} &   \mathbf{0}_{m_{\rm{tot}} \times m_{0}} \\
 \mathbf{0}_{m_{0} \times m_{\rm{tot}}} & \mathbf{I}_{m_{0}} \\
 \end{bmatrix},
\end{align}
\label{eq:chGaussianMIMO}
\end{subequations}
where 
\begin{itemize}
\item
$m_{0}$ is the number of antennas at the source, 
$m_{k}$ is the number of antennas at relay $k\in[1:N]$ with  $m_{\rm{tot}}:= \sum_{k=1}^N m_k$ (i.e., $m_{\rm{tot}}$ is the total number of antennas at the relays), and
$m_{N+1}$ is the number of antennas at the destination.

\item 
$\mathbf{y}:=[\mathbf{y}_{1};\ldots;\mathbf{y}_{N};\mathbf{y}_{N+1}] \in \mathbb{C}^{(m_{\rm{tot}} +m_{N+1})\times 1}$ is the vector of the received signals with $\mathbf{y}_{i} \in \mathbb{C}^{m_i \times 1}, i \in  [1:N+1]$ being the received signal at node $i$.

\item 
$\mathbf{x}:=[\mathbf{x}_{1};\ldots;\mathbf{x}_{N};\mathbf{x}_{0}] \in \mathbb{C}^{(m_{\rm{tot}} +m_{0})\times 1}$ is the vector of the transmitted signals where $\mathbf{x}_{i} \in \mathbb{C}^{m_i \times 1}, i \in  [0:N]$ is the signal transmitted by node $i$.
As opposed to Section~\ref{sec:sysModel} we indicate here the input of the source / node~0 as $\mathbf{x}_{0}$.


\item 
$\mathbf{z}:=[\mathbf{z}_{1};\ldots;\mathbf{z}_{N};\mathbf{z}_{N+1}]\in \mathbb{C}^{(m_{\rm{tot}} +m_{N+1})\times 1}$ is the jointly Gaussian noise vector which is assumed to have i.i.d. $\mathcal{N}(0,1)$ components.

\item 
$\mathbf{S}$ is the block diagonal matrix of dimension ${m_{\rm{tot}}\times m_{\rm{tot}}}$ to account for the state (either transmit or receive) of the relay antennas; in particular
\begin{align*}
\mathbf{S}:= \begin{bmatrix}
\mathbf{S}_1 & \mathbf{0}_{m_1 \times m_2} & \ldots & \mathbf{0}_{m_1 \times m_N} \\
\mathbf{0}_{m_2 \times m_1} & \mathbf{S}_2 & \ldots & \mathbf{0}_{m_2 \times m_N} \\
\vdots & \vdots & \vdots & \vdots \\
\mathbf{0}_{m_N \times m_1} & \mathbf{0}_{m_N \times m_2} & \ldots & \mathbf{S}_N \\
\end{bmatrix},
\ \mathbf{S}_i:= {\rm diag}[S_{i,1},\ldots,S_{i,m_i}]\in  [0:1]^{m_i},
\end{align*}
where 
$S_{i,j}=1$ if the $j$-th~antenna of the $i$-th~relay is transmitting and $S_{i,j}=0$ if it is receiving, with $j \in [1:m_i], \ i \in [1:N]$.
In this model the antennas of each relay can be switched independently of one another to transmit or receive mode for a total of $2^{m_{\rm{tot}}}$ possible states. If all the antennas at a given relay must be in the same operating mode then $\mathbf{S}_i:= S_i \ {\rm diag}[\mathbf{1}^T_{m_i}], \ S_i\in  [0:1], \ i \in [1:N]$.


\item 
$\mathbf{H} \in \mathbb{C}^{\left( m_{N+1}+m_{\rm{tot}} \right) \times \left( m_{0}+m_{\rm{tot}} \right) }$ is the constant, hence known to all nodes, channel matrix defined as
\begin{align}
\mathbf{H} :=
\begin{bmatrix}
\mathbf{H}_{\rm r \to r} & \mathbf{H}_{\rm s \to r} \\
\mathbf{H}_{\rm r \to d} & \mathbf{H}_{\rm s \to d} \\
\end{bmatrix},
\label{eq:channel2}
\end{align}
where:
\begin{itemize}

\item
$\mathbf{H}_{\rm r \to r} \in \mathbb{C}^{m_{\rm{tot}}\times m_{\rm{tot}}}$ is the block matrix which defines the network connections among the relays. In particular
\begin{align*}
\mathbf{H}_{\rm r \to r}:= 
\begin{bmatrix} 
\star & \mathbf{H}_{1,2} & \ldots &  \mathbf{H}_{1,N}  \\
\mathbf{H}_{2,1} & \star & \ldots &  \mathbf{H}_{2,N}  \\
\vdots & \vdots & \vdots & \vdots  \\
\mathbf{H}_{N,1} & \mathbf{H}_{N,2} & \ldots &  \star \\
\end{bmatrix},
\end{align*}
with $\mathbf{H}_{i,j} \in \mathbb{C}^{m_{i}\times m_{j}}$, $(i,j) \in [1: N]^2$, being the channel matrix from the $j$-th relay to the $i$-th relay. Notice that the matrices on the main diagonal of $\mathbf{H}_{\rm r \to r}$ do not matter for the channel capacity since the relays operate in HD mode.

\item
$\mathbf{H}_{\rm s \to r}:= \left [ \mathbf{H}_{1,0}; \mathbf{H}_{2,0}; \ldots; \mathbf{H}_{N,0} \right ] \in \mathbb{C}^{m_{\rm{tot}}\times m_{0}}$ is the matrix which contains the channel gains from the source / node~0 to the relays. In particular, $\mathbf{H}_{i,0} \in \mathbb{C}^{m_i\times m_{0}}$, $i\in [1: N]$, is the channel matrix from the source to the $i$-th relay.

\item 
$\mathbf{H}_{\rm r \to d}:= \left [ \mathbf{H}_{N+1,1}, \mathbf{H}_{N+1,2}, \ldots, \mathbf{H}_{N+1,N} \right ] \in \mathbb{C}^{m_{N+1} \times m_{\rm{tot}}}$ is the matrix which contains the channel gains from the relays to the destination. In particular, $\mathbf{H}_{N+1,i} \in \mathbb{C}^{{m_{N+1} \times m_i}}$, $i\in [1: N]$, is the channel matrix from the $i$-th relay to the destination.

\item
$\mathbf{H}_{\rm s \to d} \in \mathbb{C}^{m_{N+1} \times m_{0}}$ is the channel matrix between the source and the destination.
\end{itemize}

\end{itemize}

For single antenna nodes, i.e., $m_k=1, \ k \in[0:N+1]$, in~\cite{ourITjournal} we showed  that NNC is optimal to within
$1.96 (N+2)$~bits per channel use universally over all channel gains. The NNC strategy
uses independent inputs at the different nodes. Thus, since all the conditions in \eqref{eq:allassumptions} are satisfied, the result in Theorem \ref{thm:main} proves the existence of an (approximately) optimal simple schedule, with at most $N+1$ non-zero entries, for single-antenna Gaussian HD relay networks.
The goal of this section is to show that our framework immediately extends to Gaussian relay networks with multi-antenna nodes.
The main result of this section is:
\begin{thm}
\label{thm:multiantenna}
Under the assumption of independent noises, the  cut-set upper bound for the MIMO Gaussian HD network with $N$~relays
can be attained to within $1.96$~bits per channel use per antenna universally over all channel gains with NNC.
Moreover, the (approximately) optimal schedule has at most $N+1$ non-zero entries, independently on the total number of antennas in the network.
\end{thm}

\begin{IEEEproof}
To prove the constant gap 
we proceed similarly to~\cite{ourITjournal}, where the different nodes were assumed to be equipped with a single antenna. 
In particular, the main step consists of evaluating the NNC and the cut-set bounds for a general multicast Gaussian network with $K$ nodes, where each node is equipped with multiple antennas and operates in HD mode.
The derivation of the gap for the multicast scenario is reported in Appendix \ref{app:Multicast} for completeness.
Since the unicast Gaussian HD multi-relay network 
is a particular case of the multicast scenario treated in Appendix \ref{app:Multicast}, the claim follows straightforwardly.

Since all the conditions in \eqref{eq:allassumptions} are satisfied, 
Theorem~\ref{thm:main} applies. In particular, we must solve  $\max_{\mathbb{P}_{\mathbf{S}} : \mathbf{S} \in[0:1]^{{m_{\rm{tot}} }}} \min_{\mathcal{A} \subseteq [1:N]} I_{\mathcal{A}}^{(\text{fix})}$. Since what dictates the number of active states is related to the minimization over $\mathcal{A} \subseteq [1:N]$ (and not to the maximization over  $\mathbf{S} \in[0:1]^{{m_{\rm{tot}}}}$) we conclude that the optimal schedule has at most $N+1$ active states regardless of the total number of states given by ${2^{m_{\rm{tot}}}}$.
%
%
\end{IEEEproof}

\begin{figure}
\centering
\includegraphics[width=0.7\textwidth]{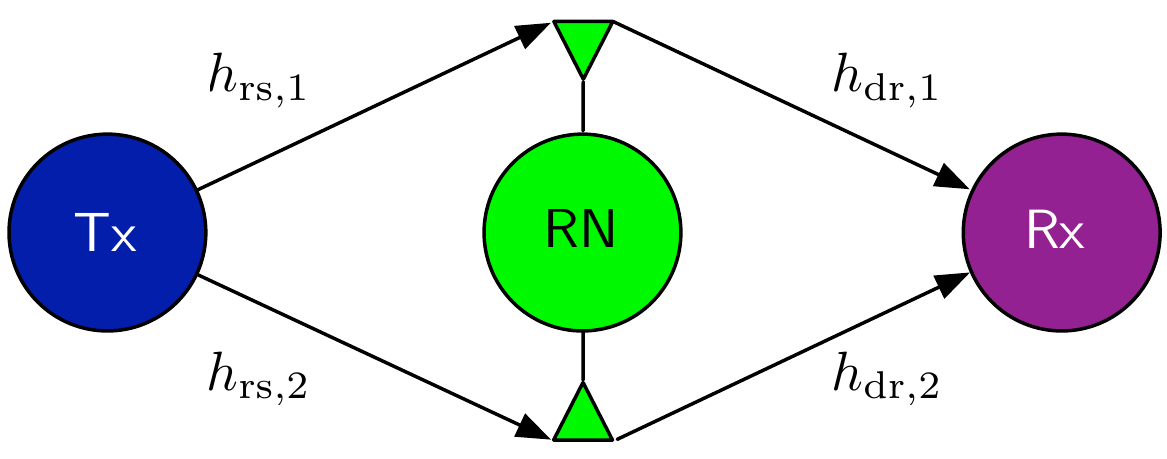}
\caption{Line network with one relay with $m_{\rm{r}}=2$ antennas, and single-antenna source and destination.}
\label{fig:ExLineNet2Ant}
\end{figure}

\subsection{Line Network Example} 
The network in Fig.~\ref{fig:ExLineNet2Ant} consists of a single-antenna source ($\mathsf{Tx}$), a single-antenna destination ($\mathsf{Rx}$) and $N=1$ relay ($\mathsf{RN}$) equipped with $m_{{\rm{r}}}=2$ antennas. Since there is no direct link between the source and the destination, this is a {\it line network}, which in the case of one relay is also a {\it diamond network.} In~\cite[Theorem~3]{OurITSingleRelay} we showed that the cut-set bound is tight for this line network with independent noises and is achieved by partial DF.
The input-output relationship is
\begin{subequations}
\begin{align}
\mathbf{y}_{{\rm{r}}} &= 
\begin{bmatrix}
(1-{S}_{1})h_{{\rm{rs}},1} \\ 
(1-{S}_{2})h_{{\rm{rs}},2} \\
\end{bmatrix} x_{0} + \mathbf{z}_{{\rm{r}}},
\\ y_{\rm{d}} &= 
\begin{bmatrix} h_{{\rm{dr}},1} & h_{{\rm{dr}},2} \\ \end{bmatrix} 
\begin{bmatrix}{S}_{1} x_{1}\\ {S}_{2} x_{2} \\ \end{bmatrix} 
+ z_{\rm{d}},
\end{align}
\label{eq:ExLineNet2Ant}
\end{subequations}
where we let (note the slightly different use of the subscripts in this section compared to the rest of the paper):
\begin{itemize}
\item
$x_{0}$ and $\mathbf{x}_{\rm{r}}=[x_{1}; \ x_{2}]$ be the signals transmitted by the source and the relay, respectively;
\item
$\mathbf{y}_{\rm{r}}=[y_{1}; \ y_{2}]$ and $y_{\rm{d}}$ be the signals received at the relay and destination, respectively; 
\item
$\mathbf{z}_{{\rm{r}}}=[z_{1}; \ z_{2}]$ and $z_{\rm{d}}$ be the noises at the relay and destination, respectively;
\item
$\mathbf{s}_{\rm{r}}=[{S}_{1}; \ {S}_{2}]$ be the state of the relay antennas;
in the following we will consider two different possible strategies at the relay: 
(i) $\mathbf{s}_{\rm{r}} \in[0:1]^2$ (i.e., the $m_{{\rm r}}=2$ antennas at the relay are switched independently of one another)
and
(ii) $\mathbf{s}_{\rm{r}}={S} \mathbf{1}_2 : {S} \in[0:1]$ (i.e., the $m_{{\rm r}}=2$ antennas at the relay are used for the same purpose);
clearly the highest rate can be attained in case~(i) since case~(ii) is a special case of case~(i) when we enforce $\mathbb{P}[{S}_{1} \not= {S}_{2}] = 0$; 
\item
the channel gains are constant and known to all nodes;
\item
the inputs are subject to the power constraints
\begin{subequations}
\begin{align}
\mathbb{E}[|x_{0}|^2]
  &=\sum_{s  \in  [0:1]^2} \lambda_s \mathbb{E}[|x_{0}|^2 | \mathbf{s}_{\rm{r}}=s] \notag
\\&=\sum_{s  \in  [0:1]^2} \lambda_s P_{0 | s} \leq 1,
\\
\mathbb{E} \left[ \| \mathbf{x}_{\rm{r}} \|^2 \right] 
  &= \Tr\left[  \sum_{s  \in  [0:1]^2} \lambda_s \mathbb{E} \left[ \mathbf{x}_{\rm{r}} \mathbf{x}^{\dagger}_{\rm{r}}  | \mathbf{s}_{\rm{r}}=s \right]  \right]\notag
\\&=\Tr\left[  \sum_{s  \in  [0:1]^2} \lambda_s  
\begin{bmatrix} 
P_{1|s} & \rho_{s} \sqrt{P_{1|s}P_{2|s}} \\ 
\rho_{s}^*\sqrt{P_{1|s}P_{2|s}} &P_{2|s} \\
\end{bmatrix} \right] \leq 1,
\end{align}
\label{eq:powpow}
\end{subequations}
where $\rho_{s} : |\rho_{s}| \in [0,1]$ is the correlation coefficient among the relay antennas in state $s  \in  [0:1]^2$. 
\end{itemize}

We start by analyzing case~(i), in which the $m_{{\rm r}}=2$ antennas at the relay are switched independently of one another.
In this network there are two cuts to consider for $I_{\mathcal{A}}^{(\text{fix})}$ in~\eqref{eq:IAfixed}, namely, $\mathcal{A}=\emptyset$ and $\mathcal{A}=\{1\}$. Recall that it suffices to evaluate $I_{\mathcal{A}}^{(\text{fix})}$ for $x_{0}$ independent of $ \mathbf{x}_{\rm{r}}$; actually, in absence of a direct source-destination link it is optimal in the cut-set bound to use $x_{0}$ independent of $ \mathbf{x}_{\rm{r}}$. Note that Gaussian inputs are not optimal in general for Gaussian networks with HD relays because information can be conveyed to the destination through random switching between listen and transmit states at the relays. To within a constant gap a fixed switching between listen and transmit states is optimal; in this case, for each state a Gaussian input is optimal. Therefore it is optimal to consider Gaussian inputs when evaluating $I_{\mathcal{A}}^{(\text{fix})}$. Moreover, from the mutual information expressions in the following, it will become clear that an optimal choice of the correlation coefficients is
$\rho_{00}=\rho_{01}=\rho_{10}=0$ and $\rho_{11} = \eu^{\jj\angle{(h_{{\rm{dr}},1}^* h_{{\rm{dr}},2}})}$.
We have
\begin{align}
I_{\emptyset}^{(\text{fix})}
  &= \sum_{s  \in  [0:1]^2} \lambda_s I \left( x_{0}, \mathbf{x}_{\rm{r}}; y_{\rm{d}} | \mathbf{s}_{\rm{r}}=s \right) \nonumber
\\&=\lambda_{00} I \left( x_{0}       ;y_{\rm{d}}|{S}_{1}=0,{S}_{2}=0\right)\nonumber
\\&+\lambda_{01} I \left( x_{0}, x_{2};y_{\rm{d}}|{S}_{1}=0,{S}_{2}=1\right)\nonumber
\\&+\lambda_{10} I \left( x_{0}, x_{1};y_{\rm{d}}|{S}_{1}=1,{S}_{2}=0\right)\nonumber
\\&+\lambda_{11} I \left( x_{0}, x_{1}, x_{2};y_{\rm{d}}|{S}_{1}=1,{S}_{2}=1\right)\nonumber
%
%
\\&=\lambda_{00} \log \left(1+0\right)\nonumber
\\&+\lambda_{01} \log \left(1+|h_{{\rm{dr}},2}|^2P_{2|01}\right)\nonumber
\\&+\lambda_{10} \log \left(1+|h_{{\rm{dr}},1}|^2P_{1|10}\right)\nonumber
\\&+\lambda_{11} \log \left(1+\left(\sqrt{|h_{{\rm{dr}},1}|^2P_{1|11}}+\sqrt{|h_{{\rm{dr}},2}|^2P_{2|11}}\right)^2
\right),
\label{eq:I0fixlinenet}
\end{align}
and
\begin{align}
I_{\{1\}}^{(\text{fix})}
  &= \sum_{s  \in  [0:1]^2} \lambda_s I \left( x_{0}; y_{\rm{d}}, \mathbf{y}_{\rm{r}} | \mathbf{x}_{\rm{r}}, \mathbf{s}_{\rm{r}}=s \right)\nonumber
\\&=\lambda_{00} I \left( x_{0}; y_{\rm{d}}, y_{1}, y_{2} | \mathbf{x}_{\rm{r}}, {S}_{1}=0,{S}_{2}=0 \right)\nonumber
\\&+\lambda_{01} I \left( x_{0}; y_{\rm{d}}, y_{1}        | \mathbf{x}_{\rm{r}}, {S}_{1}=0,{S}_{2}=1 \right)\nonumber
\\&+\lambda_{10} I \left( x_{0}; y_{\rm{d}},        y_{2} | \mathbf{x}_{\rm{r}}, {S}_{1}=1,{S}_{2}=0 \right)\nonumber
\\&+\lambda_{11} I \left( x_{0}; y_{\rm{d}}               | \mathbf{x}_{\rm{r}}, {S}_{1}=1,{S}_{2}=1 \right)\nonumber
%
%
%
\\&=\lambda_{00} \log \left(1+(|h_{{\rm{rs}},1}|^2+|h_{{\rm{rs}},2}|^2)P_{0|00} \right)\nonumber
\\&+\lambda_{01} \log \left(1+|h_{{\rm{rs}},1}|^2P_{0|01} \right)\nonumber
\\&+\lambda_{10} \log \left(1+|h_{{\rm{rs}},2}|^2P_{0|10} \right)\nonumber
\\&+\lambda_{11} \log \left(1+0 \right).
\label{eq:I1fixlinenet}
\end{align}

To determine the NNC achievable rate it suffices to remove 
the term $I \left( \mathbf{y}_{\rm{r}}; \hat{\mathbf{y}}_{\rm{r}}|x_0,\mathbf{x}_{\rm{r}},\mathbf{s}_{\rm{r}},y_{\rm{d}} \right) = m_{\rm{r}}\log(1+1/\sigma^2)$ from $I_{\emptyset}^{(\text{fix})}$ and the term $I \left( x_0; {\mathbf{y}}_{\rm{r}}|\hat{\mathbf{y}}_{\rm{r}},  y_{\rm{d}}, \mathbf{x}_{\rm{r}},\mathbf{s}_{\rm{r}} \right)  \leq \log(1+\sigma^2)$ from $I_{\{1\}}^{(\text{fix})}$, with $\sigma^2$ being the variance of the quantization noise. In what follows we will let $\sigma^2=1$ for simplicity.

{
The expressions for $I_{\emptyset}^{(\text{fix})}$ in~\eqref{eq:I0fixlinenet} and $I_{\{1\}}^{(\text{fix})}$ in~\eqref{eq:I1fixlinenet} involve power allocation across the relay states, which makes the optimization problem $\max_{{\bf \lambda}_{\rm{vect}}} \min\{I_{\emptyset}^{(\text{fix})},I_{\{1\}}^{(\text{fix})}\}$ non-linear in ${\bf \lambda}_{\rm{vect}}$.
As pointed out in Remark \ref{rem:nolindip} (see also the assumption in item~\ref{thm:main:nolambdas} in Theorem~\ref{thm:main}), in order to apply Theorem~\ref{thm:main} we must further bound the mutual information terms so that to obtain a new optimization problem with constant powers across the relay states. In particular,
see Appendix \ref{app:constgapwothconstpow}, we have that $\mathsf{C}_\text{case~(i)}$ can be upper and lower bounded to within a constant gap by
\begin{align*}
\mathsf{C}^\prime_\text{case~(i)} &= \max_{{\bf \lambda}_{\rm{vect}}} \min\{I_{\emptyset}^{(\text{fixPower})},I_{\{1\}}^{(\text{fixPower})}\},
\\ I_{\emptyset}^{(\text{fixPower})}&:=
\lambda_{00} \log \left(1+0\right)
+\lambda_{01} \log \left(1+|h_{{\rm{dr}},2}|^2\right)
\\& +\lambda_{10} \log \left(1+|h_{{\rm{dr}},1}|^2\right)
+\lambda_{11} \log \left(1+\left(\sqrt{|h_{{\rm{dr}},1}|^2}+\sqrt{|h_{{\rm{dr}},2}|^2}\right)^2 \right ),
\\ I_{\{1\}}^{(\text{fixPower})} &:=\lambda_{00} \log \left(1+|h_{{\rm{rs}},1}|^2+|h_{{\rm{rs}},2}|^2 \right)
+\lambda_{01} \log \left(1+|h_{{\rm{rs}},1}|^2\right)
\\&+\lambda_{10} \log \left(1+|h_{{\rm{rs}},2}|^2 \right)
+\lambda_{11} \log \left(1+0 \right),
\end{align*}
where the gap is
\[
\mathsf{G}_1 + \mathsf{G}_2 \leq m_{\rm{r}}\log(2) + m_{\rm{r}}\log(2) + 3\log(2)=7~\text{bits},
\] 
where the loss $3\log(2)$ is due to a fixed power allocation (see Appendix \ref{app:constgapwothconstpow}).
Now, by applying Theorem \ref{thm:multiantenna}, $\mathsf{C}^\prime_\text{case~(i)}$ (which can be straightforwardly cast into a LP as in \eqref{eq:polmnuytf}) has at most $N+1=2$ active states.

For case~(ii) (i.e., the $m_{\rm{r}}=2$ antennas at the relay are used for the same purpose), it suffices to set  $\lambda_{01}=\lambda_{10}=0$ in case~(i), i.e., to let $\lambda_{00}=1-\lambda_{11}=\lambda\in[0,1]$. With this we get that the rate in~\eqref{eq:capinidinputs} (within again $7$~bits) is
\begin{align*}
\mathsf{C}^\prime_\text{case~(ii)} &= \max_{\lambda\in[0,1]} \min\left\{ \lambda \log \left(1\!+\!|h_{{\rm{rs}},1}|^2\!+\!|h_{{\rm{rs}},2}|^2 \right), \left( 1\!-\!\lambda \right) \log \left(1+\left(\sqrt{|h_{{\rm{dr}},1}|^2}\!+\!\sqrt{|h_{{\rm{dr}},2}|^2}\right)^2 \right )  \right \}
\\& = \frac{\log \left(1+|h_{{\rm{rs}},1}|^2+|h_{{\rm{rs}},2}|^2 \right)\log \left(1+\left(\sqrt{|h_{{\rm{dr}},1}|^2}+\sqrt{|h_{{\rm{dr}},2}|^2}\right)^2 \right )}
{\log \left(1+|h_{{\rm{rs}},1}|^2+|h_{{\rm{rs}},2}|^2 \right)+\log \left(1+\left(\sqrt{|h_{{\rm{dr}},1}|^2}+\sqrt{|h_{{\rm{dr}},2}|^2}\right)^2 \right )},
\end{align*}
where the last equality follows by equating the two expressions within the min in order to find the optimal $\lambda$, which is given by
\begin{align*}
\lambda_\text{case~(ii)}= \frac{\log \left(1+\left(\sqrt{|h_{{\rm{dr}},1}|^2}+\sqrt{|h_{{\rm{dr}},2}|^2}\right)^2 \right )}{\log \left(1+|h_{{\rm{rs}},1}|^2+|h_{{\rm{rs}},2}|^2 \right)+\log \left(1+\left(\sqrt{|h_{{\rm{dr}},1}|^2}+\sqrt{|h_{{\rm{dr}},2}|^2}\right)^2 \right )}.
\end{align*}

We now show through two simple examples that not only $\mathsf{C}^\prime_\text{case~(i)} \geq \mathsf{C}^\prime_\text{case~(ii)}$, i.e., independently switching the antennas at the relay brings achievable rate gains compared to using the antennas for the same purpose, but that the difference between the two can be {\it unbounded}. In other words, at high SNR $\mathsf{C}^\prime_\text{case~(i)}$ and $\mathsf{C}^\prime_\text{case~(ii)}$ have different pre-logs / multiplexing gains / degrees of freedom.

\paragraph{Example 1}
let $|h_{{\rm{rs}},2}|=|h_{{\rm{dr}},1}|=0$ and $|h_{{\rm{rs}},1}|^2 =|h_{{\rm{dr}},2}|^2= \gamma >0$ in Fig.~\ref{fig:ExLineNet2Ant}. With this choice of the channel parameters we get
\begin{align*}
\mathsf{C}^\prime_\text{case~(i)} &
=\max_{{\bf \lambda}_{\rm{vect}}} \min \left \{ \lambda_{01} \log \left( 1 + \gamma \right)+\lambda_{11} \log \left( 1 + \gamma\right), \right.
\\& \left. \qquad \lambda_{00} \log \left( 1 + \gamma \right)+\lambda_{01} \log \left( 1 + \gamma \right) \right \}
\\& =\log \left( 1+ \gamma \right),
\end{align*}
where the last equality follows since the optimal choice of ${\bf \lambda}_{\rm{vect}}$ is given by $\lambda_{00}=\lambda_{10}=\lambda_{11}=0$ and $\lambda_{01}=1$, i.e., there is $1<N+1=2$ active state. 
For $\mathsf{C}^\prime_\text{case~(ii)}$ the optimal $\lambda$ is $1/2$ and
\begin{align*}
\mathsf{C}^\prime_\text{case~(ii)} &= \frac{\log \left( 1+\gamma \right)}{2}.
\end{align*}
It hence follows that $\mathsf{C}^\prime_\text{case~(i)} > \mathsf{C}^\prime_\text{case~(ii)}, \forall \gamma >0$. 

Moreover, the pre-log factor for $\mathsf{C}^\prime_\text{case~(i)}$ is twice that of $\mathsf{C}^\prime_\text{case~(ii)}$.
This can be interpreted as follows. 
By independently switching the $m_{\rm{r}}=2$ antennas at the relay, the achievable rate $\mathsf{C}^\prime_\text{case~(i)}$ equals (to within a constant gap) the capacity of a single-antenna relay channel with a FD relay with the source-relay and relay-destination channel gains of strength equal to $\gamma$.
On the other hand, by using the $m_{\rm{r}}=2$ antennas for the same purpose, the achievable rate $\mathsf{C}^\prime_\text{case~(ii)}$ reduces to the capacity of a single-antenna HD relay channel. 
This simple example highlights the importance of smartly switching the relay antennas in order to fully exploit the available system resources.

\paragraph{Example 2}
let $|h_{{\rm{rs}},1}|^2 =|h_{{\rm{rs}},2}|^2=|h_{{\rm{dr}},1}|^2=|h_{{\rm{dr}},2}|^2= \gamma >0$ in Fig.~\ref{fig:ExLineNet2Ant}. With this choice of the channel parameters we get
\begin{align}
\mathsf{C}^\prime_\text{case~(i)} &
=\max_{{\bf \lambda}_{\rm{vect}}} \min \left \{ \lambda_{01} \log \left( 1 + \gamma \right)+\lambda_{10} \log \left( 1 + \gamma \right)+\lambda_{11} \log \left( 1 + 4\gamma\right), \right. \nonumber
\\& \left. \qquad \lambda_{00} \log \left( 1 + 2\gamma \right)+\lambda_{01} \log \left( 1 + \gamma \right)+\lambda_{10} \log \left( 1 + \gamma \right) \right \} \nonumber
\\ & \stackrel{{\rm{(a)}}}= \max \left \{ \log \left( 1+\gamma \right), \frac{\log \left( 1+2 \gamma\right) \log \left( 1+4\gamma\right)}{\log \left( 1+2\gamma\right)+\log \left( 1+4\gamma\right)} \right \} \nonumber
\\&\stackrel{{\rm{(b)}}}= \left \{ \begin{array}{ll}
\log \left( 1+\gamma \right) & \text{if} \ \gamma \geq 0.752
\\ \frac{\log \left( 1+2 \gamma\right) \log \left( 1+4\gamma\right)}{\log \left( 1+2\gamma\right)+\log \left( 1+4\gamma\right)} & \text{otherwise}
\end{array},
\right.
\label{eq:Ccaseibound}
\end{align}
where the equality in (a) follows since among the ten possible (approximately) optimal simple schedules ${\bf \lambda}_{\rm{vect}}$ (six possible ${\bf \lambda}_{\rm{vect}}$ with two active states plus four possible ${\bf \lambda}_{\rm{vect}}$ with one active state), it is easy to see that only the two cases ${\bf \lambda}_{\rm{vect}}=[0,0,1,0]$ and ${\bf \lambda}_{\rm{vect}}=[\lambda,0,0,1-\lambda]$, with $\lambda=\frac{\log \left( 1+4 \gamma\right) }{\log \left( 1+2\gamma\right)+\log \left( 1+4\gamma\right)}$, have to be considered and the equality in (b) follows from numerical evaluations. Thus, if $\gamma \geq 0.752$ the (approximately) optimal schedule has $1<N+1=2$ active state (i.e., $\lambda_{10}$ only), otherwise it has $N+1=2$ active states (i.e., $\lambda_{00}$ and $\lambda_{11}$).

For $\mathsf{C}^\prime_\text{case~(ii)}$ we obtain that the optimal $\lambda=\frac{\log \left( 1+4 \gamma\right) }{\log \left( 1+2\gamma\right)+\log \left( 1+4\gamma\right)}$ and
\begin{align}
\mathsf{C}^\prime_\text{case~(ii)} &= \frac{\log \left( 1+2 \gamma\right) \log \left( 1+4\gamma\right)}{\log \left( 1+2\gamma\right)+\log \left( 1+4\gamma\right)}.
\label{eq:Ccaseiibound}
\end{align}
It hence follows that $\mathsf{C}^\prime_\text{case~(i)} > \mathsf{C}^\prime_\text{case~(ii)}, \forall \gamma \geq 0.752$, as can also be observed from Fig. \ref{fig:WFversusBounding} (blue dashed line for $\mathsf{C}^\prime_\text{case~(i)}$ versus red red dashed line for $\mathsf{C}^\prime_\text{case~(ii)}$). 

\begin{figure}
\centering
\includegraphics[width=0.7\textwidth]{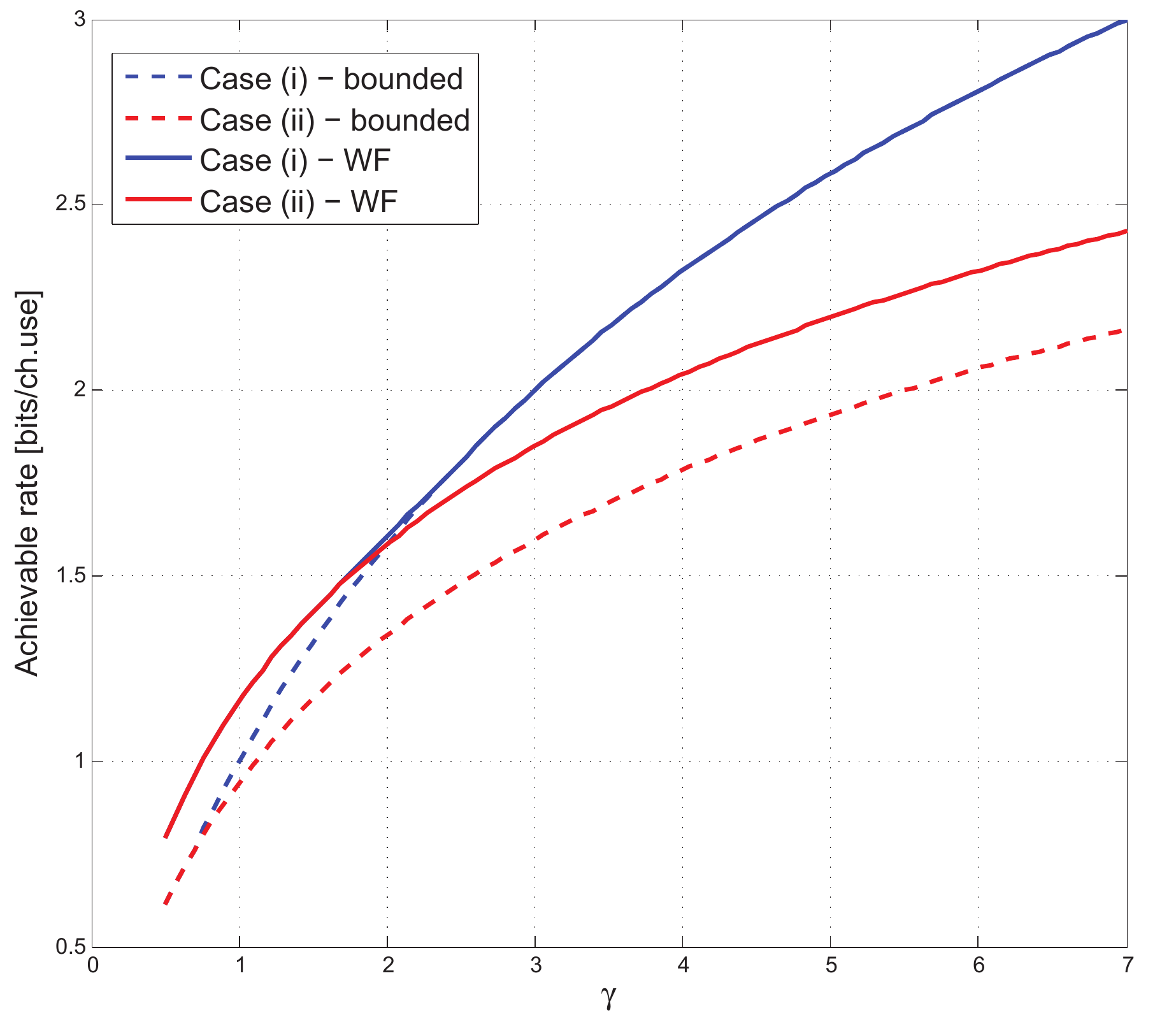}
\caption{$\mathsf{C}^\prime_\text{case~(i)}$,$\mathsf{C}^\prime_\text{case~(ii)}$,$\mathsf{C}^{\prime\prime}_\text{case~(i)}$,
$\mathsf{C}^{\prime\prime}_\text{case~(ii)}$ versus different values of $\gamma$.}
\label{fig:WFversusBounding}
\end{figure}

Fig. \ref{fig:WFversusBounding} also shows the achievable rates $\mathsf{C}^{\prime\prime}_\text{case~(i)} = \max_{{\bf \lambda}_{\rm{vect}}} \min\{I_{\emptyset}^{(\text{fix})},I_{\{1\}}^{(\text{fix})}\}$ (solid blue line) and $\mathsf{C}^{\prime\prime}_\text{case~(ii)}$ (solid red line) obtained by optimizing the powers in $I_{\emptyset}^{(\text{fix})}$ in \eqref{eq:I0fixlinenet} and $I_{\{1\}}^{(\text{fix})}$ in \eqref{eq:I1fixlinenet} across the different states by Water Filling (WF), as described in Appendix \ref{app:waterfilling}. In particular, under the channel conditions considered in this example, from Appendix \ref{app:waterfilling} we get that the optimal power allocation can be found by solving 
\begin{align*}
\mathsf{C}^{\prime\prime}_\text{case~(i)} &= \max_{\lambda\in[0,1],\nu \geq 0}\left\{
\lambda \log^+ \left(\gamma \nu \right) +\frac{1-\lambda}{2}\log^+ \left(2\gamma \nu \right)
\right\}
\\
\nu &: \ \lambda \left(\nu-\frac{1}{\gamma}\right)^+ + \frac{1-\lambda}{2} \left(\nu-\frac{1}{2\gamma} \right)^+ = 1,
\end{align*}
where $\lambda_{01}+\lambda_{10} = \lambda \in[0,1], \ \lambda_{00} = \lambda_{11} = \frac{1-\lambda}{2}$,
which is equal to
\begin{align}
\mathsf{C}^{\prime\prime}_\text{case~(i)} &= \max_{\lambda\in[0,1]}\left\{
\lambda \log \left( \frac{3 \lambda +1 }{2 (\lambda +1)} + \frac{2}{\lambda +1} \gamma \right) +\frac{1-\lambda}{2}\log \left( \frac{3 \lambda +1 }{\lambda +1} + \frac{4}{\lambda +1} \gamma \right)
\right\},
\label{eq:CcaseiWF}
\end{align}
which is represented by the blue solid line in Fig. \ref{fig:WFversusBounding}. For case~(ii) it suffices to set $\lambda=0$ in $\mathsf{C}^{\prime\prime}_\text{case~(i)} $; with this we obtain 
\begin{align}
\mathsf{C}^{\prime\prime}_\text{case~(ii)}&= \frac{1}{2} \log \left( 1+4\gamma\right),
\label{eq:CcaseiiWF}
\end{align}
which is represented by the red solid line in Fig. \ref{fig:WFversusBounding}.

From Fig. \ref{fig:WFversusBounding} we observe that the highest rates are achieved by optimizing the powers across the different states (solid lines versus dashed lines). However, as also highlighted in Remark \ref{rem:nolindip} (see also the assumption in item~\ref{thm:main:nolambdas} in Theorem~\ref{thm:main}), with optimal power allocation there are no guarantees that the (approximately) optimal schedule is simple. This is exactly what we observe in this example for which the optimal $\lambda \in [0,1]$ that maximizes $\mathsf{C}^{\prime\prime}_\text{case~(i)}$ in \eqref{eq:CcaseiWF} is neither zero nor one, i.e., the  schedule has $3>N+1=2$ active states. 
From Fig. \ref{fig:WFversusBounding} we also notice that the difference between the solid lines (obtained by optimizing the powers across the states) and the dashed lines (obtained with a constant / fixed power allocation) is at most $0.1977$~bits for case~(i) (blue lines) and $0.2636$~bits for case~(ii) (red lines). These differences are far smaller than the  $3$~bits computed analytically in Appendix \ref{app:constgapwothconstpow}, showing that the theoretical gap of $3$~bits is very conservative, at least for this choice of the channel parameters.

}

\section{Conclusions}
\label{sec:Concl}
In this work we studied networks with $N$ half-duplex relays. 
For such networks, the capacity must be optimized over the $2^N$ possible listen-transmit relay configurations.
We proved that, if the noises are independent and independent inputs are approximately optimal in the cut-set bound, then the approximately optimal schedule is {\it simple} in the sense that at most $N+1$ relay configurations have a non-zero probability.
We proposed a convergent iterative polynomial-time algorithm to find the (approximately) optimal simple schedule.

We applied the result to Gaussian noise networks with multi-antenna nodes, where the antennas at the relays can be switched between listen and transmit state independently of one another. We showed that the cut-set outer bound can be achieved to within a constant gap (which depends on the total number of antennas but not on the channel gains) and that the corresponding optimal schedule is simple, i.e., the number of active states only depends on the number of relays. 
Through a line-network example we showed that independently switching the antennas at each relay can provide a strictly larger pre-log / multiplexing gain compared to using the antennas for the same purpose.

\appendices

\section{Proof that $I_{\mathcal{A}}^{(\text{fix})}$ in~\eqref{eq:IAfixed} is submodular}
\label{app:submodularityProof}

Consider two possible cuts of the network represented by 
$\mathcal{A}_1, \mathcal{A}_2 \subseteq [1:N]$ and let 
\begin{align*}
\begin{array}{ll}
B_0 := \mathcal{A}_1 \cap \mathcal{A}_2, &
B_1 := \mathcal{A}_1\backslash \mathcal{A}_2, \\
B_2 := \mathcal{A}_2\backslash \mathcal{A}_1, &
B_3 := [1:N]\backslash(\mathcal{A}_1\cup \mathcal{A}_2),
\end{array}
\end{align*}
so that, $B_j, j\in[0:3]$ is a partition of $[1:N]$ and thus 
\begin{align*}
\begin{array}{ll}
\mathcal{A}_1 =B_0 \cup B_1, &
\mathcal{A}_2 =B_0 \cup B_2, \\
\mathcal{A}_1 \cap \mathcal{A}_2 = B_0, &
[1:N]\backslash (\mathcal{A}_1 \cup \mathcal{A}_2) = B_3.
\end{array}
\end{align*}
Let $X_{\mathcal{A}} := \{X_i : i\in \mathcal{A}\}$ and
$X_{(n)} := \{X_i : i\in B_n\}, \ n\in[0:3]$.
We write
$I_{\mathcal{A}}^{(\text{fix})} 
  = H \left(Y_{N+1},Y_{\mathcal{A}}|X_{\mathcal{A}},S_{[1:N]} \right)
  - H \left( Y_{N+1},Y_{\mathcal{A}}  | X_{[1:N+1]},S_{[1:N]} \right)
$. We next show that, under the assumption of ``independent noises'' in~\eqref{eq:indipnoises}, the function $h_1\left( \mathcal{A}\right):=H \left( Y_{N+1},Y_{\mathcal{A}}  | X_{[1:N+1]},S_{[1:N]} \right)$ is modular and that, under the assumption of independent inputs in~\eqref{eq:indipinputs}, the function $h_2\left( \mathcal{A}\right):=H \left(Y_{N+1},Y_{\mathcal{A}}|X_{\mathcal{A}},S_{[1:N]} \right)$ is submodular; these two facts imply that $I_{\mathcal{A}}^{(\text{fix})}$ in \eqref{eq:IAfixed} is submodular.

For $h_1\left( \mathcal{A}\right)$ we have
\begin{align*}
&h_1 \left( \mathcal{A}_1\right)+h_1 \left( \mathcal{A}_2\right) -h_1\left( \mathcal{A}_1 \cup \mathcal{A}_2 \right)- h_1 \left( \mathcal{A}_1 \cap \mathcal{A}_2 \right)
\\& = H \left( Y_{N+1},Y_{(0)},Y_{(1)} | X_{[1:N+1]},S_{[1:N]} \right) + H \left( Y_{N+1},Y_{(0)},Y_{(2)} | X_{[1:N+1]},S_{[1:N]} \right)
\\& \quad - H \left( Y_{N+1},Y_{(0)},Y_{(1)},Y_{(2)} | X_{[1:N+1]},S_{[1:N]} \right)
-H \left( Y_{N+1},Y_{(0)}| X_{[1:N+1]},S_{[1:N]} \right)
\\& = H \left( Y_{(1)} | Y_{N+1},Y_{(0)},X_{[1:N+1]},S_{[1:N]} \right) + H \left( Y_{(2)} | Y_{N+1},Y_{(0)},X_{[1:N+1]},S_{[1:N]} \right)
\\& \quad - H \left( Y_{(1)},Y_{(2)} | Y_{N+1},Y_{(0)},X_{[1:N+1]},S_{[1:N]} \right)
\\&= I \left( Y_{(1)};Y_{(2)} | Y_{N+1},Y_{(0)},X_{[1:N+1]},S_{[1:N]}\right)=0,
\end{align*}
where the last equality follows because of the assumption of ``independent noises'' in~\eqref{eq:indipnoises}. 
Therefore $h_1\left( \mathcal{A}\right)$  is modular.

For $h_2\left( \mathcal{A}\right)$ we have
\begin{align*}
&h_2 \left( \mathcal{A}_1\right)+h_2 \left( \mathcal{A}_2\right) -h_2\left( \mathcal{A}_1 \cup \mathcal{A}_2 \right)- h_2 \left( \mathcal{A}_1 \cap \mathcal{A}_2 \right)
\\&=H \left (Y_{N+1},Y_{(0)},Y_{(1)}|X_{(0)},X_{(1)},S_{[1:N]} \right )
  +H(Y_{N+1},Y_{(0)},Y_{(2)}|X_{(0)},X_{(2)},S_{[1:N]})
\\& \quad -H(Y_{N+1},Y_{(0)},Y_{(1)},Y_{(2)}|X_{(0)},X_{(1)},X_{(2)},S_{[1:N]})
 - H(Y_{N+1},Y_{(0)}|X_{(0)},S_{[1:N]})
\\&=H(Y_{N+1},Y_{(0)}|X_{(1)},S_{[1:N]},X_{(0)})
 + H(Y_{N+1},Y_{(0)}|X_{(2)},S_{[1:N]},X_{(0)})
\\&  \quad - H(Y_{N+1},Y_{(0)}|X_{(1)},X_{(2)},S_{[1:N]},X_{(0)})
  - H(Y_{N+1},Y_{(0)}|S_{[1:N]},X_{(0)})
\\& \quad + H(Y_{(1)}|X_{(1)},S_{[1:N]}, Y_{N+1},X_{(0)},Y_{(0)})  
   + H(Y_{(2)}|X_{(2)},S_{[1:N]}, Y_{N+1},X_{(0)},Y_{(0)})
\\& \quad - H(Y_{(1)},Y_{(2)}|X_{(1)},X_{(2)},S_{[1:N]}, Y_{N+1},X_{(0)},Y_{(0)})
\\&=I(Y_{N+1},Y_{(0)};X_{(2)}|X_{(1)},S_{[1:N]},X_{(0)}) 
- I(Y_{N+1},Y_{(0)}; X_{(2)}|S_{[1:N]},X_{(0)})
\\& \quad + I(Y_{(1)}; X_{(2)}|X_{(1)},S_{[1:N]}, Y_{N+1},X_{(0)},Y_{(0)})  
    + I(Y_{(2)}; Y_{(1)},X_{(1)}|X_{(2)},S_{[1:N]}, Y_{N+1},X_{(0)},Y_{(0)})
\\&=I(X_{(1)};X_{(2)}|S_{[1:N]},X_{(0)}, Y_{N+1},Y_{(0)})+ I(Y_{(1)}; X_{(2)}|X_{(1)},S_{[1:N]}, Y_{N+1},X_{(0)},Y_{(0)})  
\\& \quad {- I(X_{(1)}; X_{(2)}|S_{[1:N]},X_{(0)})}
 + I(Y_{(2)}; Y_{(1)},X_{(1)}|X_{(2)},S_{[1:N]}, Y_{N+1},X_{(0)},Y_{(0)})
\geq 0,
\end{align*}
where the last inequality follows because the ``independent inputs'' assumption in~\eqref{eq:indipinputs} implies ${I(X_{(1)}; X_{(2)}|S_{[1:N]},X_{(0)}) =0}$. This shows that $h_2\left( \mathcal{A}\right)$  is submodular.

\section{Gap result for Gaussian multicast networks with multi-antenna nodes}
\label{app:Multicast}
A Gaussian multicast network with $K$ nodes, each equipped with $m_k, k \in [1:K]$ antennas, is defined similarly to the Gaussian multi-relay network in \eqref{eq:chGaussianMIMO}, except that now each node $k\in[1:K]$, with channel input $(\mathbf{x}_k,\mathbf{s}_k)$ and channel output $\mathbf{y}_k$, has an independent message of rate $R_k$ to be decoded by the nodes indexed by $\mathcal{D}\subseteq[1:K]$. 
The channel input/output relationship of this HD Gaussian multicast network reads 
\begin{align*}
\mathbf{y} &= (\mathbf{I}_{M_{\rm{tot}}} - \mathbf{S})  \mathbf{H}  \mathbf{S}  \mathbf{x}+\mathbf{z}
\\& = \underbrace{\begin{bmatrix}
\star & \left( \mathbf{I}_{m_1} - \mathbf{S}_1 \right) \mathbf{H}_{1,2} \mathbf{S}_2 & \ldots & \left( \mathbf{I}_{m_1} - \mathbf{S}_1 \right) \mathbf{H}_{1,K} \mathbf{S}_K \\
\left( \mathbf{I}_{m_2} - \mathbf{S}_2 \right) \mathbf{H}_{2,1} \mathbf{S}_1 & \star & \ldots & \left( \mathbf{I}_{m_2} - \mathbf{S}_2 \right) \mathbf{H}_{2,K} \mathbf{S}_K \\
\vdots & \vdots & \vdots & \vdots \\
\left( \mathbf{I}_{m_K} - \mathbf{S}_K \right) \mathbf{H}_{K,1} \mathbf{S}_1 & \left( \mathbf{I}_{m_K} - \mathbf{S}_K \right) \mathbf{H}_{K,2} \mathbf{S}_2 & \ldots & \star \\
\end{bmatrix}}_{\mathbf{H}_{{\rm{tot}}}} \mathbf{x}+\mathbf{z},
\end{align*}
with $M_{\rm{tot}}:= \sum_{k=1}^K m_k$. We let $\mathcal{C}_{\rm multicast}$ be the capacity region. By following similar bounding steps as in~\cite[${\rm{eq.}}(27)$]{ourITjournal} and by keeping in mind that each node $k \in [1:K]$ is now equipped with $m_k$ antennas, we have that NNC achieves the following rate region
\begin{align}
\mathcal{C}_{\rm multicast} \supseteq  \bigcup & \left\{
\sum_{i\in\mathcal{A}}R_i  \leq 
 {\sum_{s\in [0:1]^{M_{\rm{tot}}}} } \lambda_{s}  \log\left|\mathbf{I}_{m_{\mathcal{A}^c}} + \frac{1}{1+\sigma^2} \mathbf{H}_{\mathcal{A},s}  \mathbf{H}_{\mathcal{A},s}^H\right|-m_{\mathcal{A}} \log\left(1+\frac{1}{\sigma^2}\right)
\right.\nonumber
\\ & \left. \quad 
\quad  \text{such that} \  \mathcal{A}\subseteq[1:K], \ {\mathcal{A}\not=\emptyset}, \ \mathcal{A}^c \cap \mathcal{D} \not=\emptyset
\Bigg \} \right.,
\label{eq:MultiCastlow}
\end{align}
where $\sigma^2$ is the variance of the quantization noise which does not depend neither on the user index $k \in [1:K]$ nor on the antenna index $j \in [1:m_k]$ of user $k$ and where the matrix $\mathbf{H}_{\mathcal{A},s} \in \mathbb{C}^{m_{\mathcal{A}^c} \times m_{\mathcal{A}}}$ is defined as $\mathbf{H}_{\mathcal{A},s} := \left [ \mathbf{H}_{{\rm{tot}}} \right ]_{\mathcal{A}^c,\mathcal{A}}$, with $m_{\mathcal{A}^c}:=\sum_{i=1,i \in \mathcal{A}^c}^K m_i$ and $m_{\mathcal{A}}:=\sum_{i=1,i \in \mathcal{A}}^K m_i$.

Similarly, by proceeding as in~\cite[${\rm{eq.}}(29)$]{ourITjournal}, the cut-set upper bound can be further upper bounded as
\begin{align}
&\mathcal{C}_{\rm multicast}  \subseteq   \bigcup  \left\{
\sum_{i\in\mathcal{A}}R_i
\leq 
m_{\mathcal{A}}\log(2)
 +  \sum_{s\in[0:1]^{M_{\rm{tot}}}} \lambda_{s} \ 
\log\left|\mathbf{I}_{m_{\mathcal{A}^c}} + \frac{1}{\gamma} \ \mathbf{H}_{\mathcal{A},s} \mathbf{H}_{\mathcal{A},s}^H\right|
\right.\nonumber
\\ &\left. 
+ m_{\mathcal{A}} \
  \frac{ \log\left( \eu \max\left \{1,\frac{\gamma}{\eu} \ \frac{m_{\mathcal{A}}}{\min\{m_{\mathcal{A}},m_{\mathcal{A}^c}\}}\right \}\right) }
  {\max\left \{\frac{\eu}{\gamma},\frac{m_{\mathcal{A}}}{\min\{m_{\mathcal{A}},m_{\mathcal{A}^c}\}}\right\}}
 \ \text {such that} \ \mathcal{A}\subseteq[1:K], \ {\mathcal{A}\not=\emptyset}, \ \mathcal{A}^c \cap \mathcal{D} \not=\emptyset
\right\}.
\label{eq:MultiCastup}
\end{align}
By taking the difference between the outer bound in \eqref{eq:MultiCastup} and the lower bound in \eqref{eq:MultiCastlow} (see also~\cite[${\rm{eq.}}(30)$]{ourITjournal}), we obtain
$\mathsf{GAP} \leq 1.96 M_{\rm{tot}}$
bits per channel use.

{
\section{Upper and lower bounds for $I_{\emptyset}^{(\text{fix})}$ in \eqref{eq:I0fixlinenet} and $I_{\{1\}}^{(\text{fix})}$ in \eqref{eq:I1fixlinenet}}
\label{app:constgapwothconstpow}
In this section we prove that
\begin{align}
\mathsf{C}^\prime_\text{case~(i)} -  \log(2) \leq \mathsf{C}^{\prime\prime}_\text{case~(i)}  \leq
\mathsf{C}^\prime_\text{case~(i)} + 2 \log(2),
\label{eq:I0fixI1fixbound}
\end{align}
where
\begin{align*}
\mathsf{C}^{\prime\prime}_\text{case~(i)} &:= \max_{{\bf \lambda}_{\rm{vect}}} \min\{I_{\emptyset}^{(\text{fix})},I_{\{1\}}^{(\text{fix})}\},
\\ \mathsf{C}^\prime_\text{case~(i)} &:= \max_{{\bf \lambda}_{\rm{vect}}} \min\{I_{\emptyset}^{(\text{fixPower})},I_{\{1\}}^{(\text{fixPower})}\} ,
\\ I_{\emptyset}^{(\text{fixPower})}&:=
    \lambda_{00} \log \left(1+0\right)
   +\lambda_{01} \log \left(1+|h_{{\rm{dr}},2}|^2\right)
\\&+\lambda_{10} \log \left(1+|h_{{\rm{dr}},1}|^2\right)
   +\lambda_{11} \log \left(1+\left(\sqrt{|h_{{\rm{dr}},1}|^2}+\sqrt{|h_{{\rm{dr}},2}|^2}\right)^2 \right ),
\\ I_{\{1\}}^{(\text{fixPower})} &:=
    \lambda_{00} \log \left(1+|h_{{\rm{rs}},1}|^2+|h_{{\rm{rs}},2}|^2 \right)
   +\lambda_{01} \log \left(1+|h_{{\rm{rs}},1}|^2\right)
\\&+\lambda_{10} \log \left(1+|h_{{\rm{rs}},2}|^2 \right)
   +\lambda_{11} \log \left(1+0 \right).
\end{align*}
We start by noting that in \eqref{eq:powpow} we can assume, without loss of optimality that:
(i) $P_{0|11}=0$, since the direct link is absent, the source does not transmit when both the $m_{\rm{r}}=2$ antennas at the relay are transmitting; and that
(ii) $P_{1|00}=P_{1|01}=0$ (resp. $P_{2|00}=P_{2|10}=0$), since for the HD constraint when the first (resp. second) antenna at the relay is receiving the relay's transmit power on that antenna is zero.
With this, we let
\begin{align*}
  & P_{0|00} = \frac{\alpha_0}{\lambda_{00}}, \ P_{0|01} = \frac{\beta_0}{\lambda_{01}}, \ P_{0|10} = \frac{\gamma_0}{\lambda_{10}},
\\& P_{2|01} = \frac{\alpha_1}{\lambda_{01}}, \ P_{1|10} = \frac{\beta_1}{\lambda_{10}}, \ P_{1|11} = \frac{\gamma_1}{\lambda_{11}}, \ P_{2|11} = \frac{\delta_1}{\lambda_{11}},
\end{align*}
where $\alpha_0+\beta_0+\gamma_0 \leq 1$ and $\alpha_1+\beta_1+\gamma_1+\delta_1 \leq 1$ in order to meet the power constraints in \eqref{eq:powpow}. We now upper bound $\mathsf{C}^{\prime\prime}_\text{case~(i)} = \max_{{\bf \lambda}_{\rm{vect}}} \min\{I_{\emptyset}^{(\text{fix})},I_{\{1\}}^{(\text{fix})}\}$ as follows
\begin{align*}
\mathsf{C}^{\prime\prime}_\text{case~(i)} 
 &= \max_{{\bf \lambda}_{\rm{vect}}} \min\{I_{\emptyset}^{(\text{fix})},I_{\{1\}}^{(\text{fix})}\} 
\\& =  \max_{{\bf \lambda}_{\rm{vect}}} \min \left \{
 \lambda_{01} \log \left(1+|h_{{\rm{dr}},2}|^2\frac{\alpha_1}{\lambda_{01}}\right)
+\lambda_{10} \log \left(1+|h_{{\rm{dr}},1}|^2\frac{\beta_1 }{\lambda_{10}}\right) \right.
\\& \left. \qquad +\lambda_{11} \log \left(1+\left(\sqrt{|h_{{\rm{dr}},1}|^2\frac{\gamma_1}{\lambda_{11}}}+\sqrt{|h_{{\rm{dr}},2}|^2\frac{\delta_1}{\lambda_{11}}}\right)^2
\right), \right.
\\& \left. 
 \qquad \lambda_{00} \log \left(1+(|h_{{\rm{rs}},1}|^2+|h_{{\rm{rs}},2}|^2)\frac{\alpha_0}{\lambda_{00}}\right)
+\lambda_{01} \log \left(1+|h_{{\rm{rs}},1}|^2\frac{\beta_0}{\lambda_{01}} \right) \right.
\\& \left. \qquad+\lambda_{10} \log \left(1+|h_{{\rm{rs}},2}|^2\frac{\gamma_0}{\lambda_{10}} \right)
\right \} 
\\&\leq 
 \max_{{\bf \lambda}_{\rm{vect}}}  H({\bf \lambda}_{\rm{vect}})+\min \Big \{
 \lambda_{01} \log \left(\lambda_{01}+|h_{{\rm{dr}},2}|^2 \alpha_1\right)
 +\lambda_{10} \log \left(\lambda_{10}+|h_{{\rm{dr}},1}|^2 \beta_1\right) 
\\& \qquad  +\lambda_{11} \log \left(\lambda_{11}+\left(\sqrt{|h_{{\rm{dr}},1}|^2\gamma_1}+\sqrt{|h_{{\rm{dr}},2}|^2 \delta_1}\right)^2
\right),
\\&
\qquad \lambda_{00} \log \left(\lambda_{00}+(|h_{{\rm{rs}},1}|^2+|h_{{\rm{rs}},2}|^2)\alpha_0\right) 
+\lambda_{01} \log \left(\lambda_{01}+|h_{{\rm{rs}},1}|^2 \beta_0 \right) 
\\& \qquad +\lambda_{10} \log \left(\lambda_{10}+|h_{{\rm{rs}},2}|^2 \gamma_0 \right)
\Big\} 
\\& \leq 2\log(2) + \max_{{\bf \lambda}_{\rm{vect}}} \min \Big\{
 \lambda_{01} \log \left(1+|h_{{\rm{dr}},2}|^2\right)
+\lambda_{10} \log \left(1+|h_{{\rm{dr}},1}|^2 \right) 
\\& \qquad +\lambda_{11} \log \left(1+\left(\sqrt{|h_{{\rm{dr}},1}|^2 }+\sqrt{|h_{{\rm{dr}},2}|^2 }\right)^2 \right),
\\&
\qquad \lambda_{00} \log \left(1+|h_{{\rm{rs}},1}|^2+|h_{{\rm{rs}},2}|^2\right) 
+\lambda_{01} \log \left(1+|h_{{\rm{rs}},1}|^2  \right) 
+\lambda_{10} \log \left(1+|h_{{\rm{rs}},2}|^2  \right)
\Big \},
\end{align*}
where the two inequalities follow because:
(i) the entropy of a discrete random variable can be upper bounded by the logarithm of the size of its support 
(i.e., $H({\bf \lambda}_{\rm{vect}}) \leq \log(4)$);
(ii) by further upper bounding the power splits by setting $\alpha_{i}=\beta_{i}=\gamma_{i}=\delta_1= 1$,  $i \in [0:1]$; 
(iii) by further upper bounding all the $\lambda_{s}, s \in [0:1]^2$ inside the logarithms by one.

We now lower bound $\mathsf{C}^{\prime\prime}_\text{case~(i)} = \max_{{\bf \lambda}_{\rm{vect}}} \min\{I_{\emptyset}^{(\text{fix})},I_{\{1\}}^{(\text{fix})}\}$ as follows
\begin{align*}
\mathsf{C}^{\prime\prime}_\text{case~(i)} & = \max_{{\bf \lambda}_{\rm{vect}}} \min\{I_{\emptyset}^{(\text{fix})},I_{\{1\}}^{(\text{fix})}\} \nonumber
\\& =  \max_{{\bf \lambda}_{\rm{vect}}} \min \left \{
\lambda_{01} \log \left(1+|h_{{\rm{dr}},2}|^2 \frac{\alpha_1}{\lambda_{01}}\right)
+\lambda_{10} \log \left(1+|h_{{\rm{dr}},1}|^2\frac{\beta_1}{\lambda_{10}}\right) \right. \nonumber
\\& \left. \qquad
+\lambda_{11} \log \left(1+\left(\sqrt{|h_{{\rm{dr}},1}|^2\frac{\gamma_1}{\lambda_{11}}}+\sqrt{|h_{{\rm{dr}},2}|^2\frac{\delta_1}{\lambda_{11}}}\right)^2
\right), \right. \nonumber
\\ & \left . \qquad
\lambda_{00} \log \left(1+(|h_{{\rm{rs}},1}|^2+|h_{{\rm{rs}},2}|^2)\frac{\alpha_0}{\lambda_{00}}\right)
+\lambda_{01} \log \left(1+|h_{{\rm{rs}},1}|^2\frac{\beta_0}{\lambda_{01}} \right) \right. \nonumber
\\ & \left. \qquad
+\lambda_{10} \log \left(1+|h_{{\rm{rs}},2}|^2\frac{\gamma_0}{\lambda_{10}} \right)
\right \} \nonumber
\\& \geq  - \log(2) + \max_{{\bf \lambda}_{\rm{vect}}} \min \left \{
\lambda_{01} \log \left(1+|h_{{\rm{dr}},2}|^2 \right)
+\lambda_{10} \log \left(1+|h_{{\rm{dr}},1}|^2\right) 
\right. \nonumber
\\& \left. \qquad
+\lambda_{11} \log \left(1+\left(\sqrt{|h_{{\rm{dr}},1}|^2}+\sqrt{|h_{{\rm{dr}},2}|^2}\right)^2
\right), \right. \nonumber
\\ & \left . \qquad
\lambda_{00} \log \left(1+|h_{{\rm{rs}},1}|^2+|h_{{\rm{rs}},2}|^2\right)
\!+\!\lambda_{01} \log \left(1+|h_{{\rm{rs}},1}|^2\right) 
\!+\!\lambda_{10} \log \left(1+|h_{{\rm{rs}},2}|^2 \right)
\right \},
\end{align*}
where the inequality follows by (i) setting $\alpha_1=\lambda_{01}$, $\beta_1=\lambda_{10}$, $\gamma_1=\delta_1=\frac{\lambda_{11}}{2}$, $\alpha_0=\lambda_{00}$, $\beta_0=\lambda_{01}$ and $\gamma_0 = \lambda_{10}$ (note that with these power splits the power constraints in \eqref{eq:powpow} are satisfied), (ii) since $\log \left(1+\left( \sqrt{\frac{a}{2}} + \sqrt{\frac{c}{2}}\right)^2\right) =\log \left(1+\frac{1}{2}\left( \sqrt{a} + \sqrt{c}\right)^2\right) \geq \log \left(\frac{1}{2}+\frac{1}{2}\left( \sqrt{a} + \sqrt{c}\right)^2\right) =\log \left(1+\left( \sqrt{a} + \sqrt{c}\right)^2\right) - \log(2)$ and (iii) by removing the term $\log(2)$ also from the second term within the $\min$. 

Thus, by considering the difference between the upper 
and the lower bounds 
we obtain the result in \eqref{eq:I0fixI1fixbound}.

\section{Water filling power allocation for $I_{\emptyset}^{(\text{fix})}$ in \eqref{eq:I0fixlinenet} and $I_{\{1\}}^{(\text{fix})}$ in \eqref{eq:I1fixlinenet}}
\label{app:waterfilling}

By optimizing the powers in the different relay states subject to the power constraints in \eqref{eq:powpow} we have
\begin{align*}
\mathsf{C}^{\prime\prime}_\text{case~(i)} &= \max_{{\bf \lambda}_{\rm{vect}}} \min\{I_{\emptyset}^{(\text{fix})},I_{\{1\}}^{(\text{fix})}\},
\end{align*}
where $I_{\emptyset}^{(\text{fix})}$ and $I_{\{1\}}^{(\text{fix})}$ are defined in \eqref{eq:I0fixlinenet} and in \eqref{eq:I1fixlinenet}, respectively. By writing the Lagrangian of the optimization problem above (subject to the power constraints in \eqref{eq:powpow}) 
we obtain
\begin{align*}
I_{\emptyset}^{(\text{fix})}
  &=\lambda_{01} \log^+ \left(\nu_0|h_{{\rm{dr}},2}|^2\right)
   +\lambda_{10} \log^+ \left(\nu_0|h_{{\rm{dr}},1}|^2\right)
   +\lambda_{11} \log^+ \left(\nu_0(|h_{{\rm{dr}},1}|^2 + |h_{{\rm{dr}},2}|^2)\right)
\\& \nu_0 : 
  \lambda_{01} \left(\nu_0-\frac{1}{|h_{{\rm{dr}},2}|^2} \right)^+
 +\lambda_{10} \left(\nu_0-\frac{1}{|h_{{\rm{dr}},1}|^2} \right)^+
 +\lambda_{11} \left(\nu_0-\frac{1}{|h_{{\rm{dr}},1}|^2 + |h_{{\rm{dr}},2}|^2} \right)^+
 =1,
\\
I_{\{1\}}^{(\text{fix})}
  &=\lambda_{00} \log^+ \left(\nu_1(|h_{{\rm{rs}},1}|^2+|h_{{\rm{rs}},2}|^2) \right)
   +\lambda_{01} \log^+ \left(\nu_1|h_{{\rm{rs}},1}|^2 \right)
   +\lambda_{10} \log^+ \left(\nu_1|h_{{\rm{rs}},2}|^2 \right),
\\& \nu_1 : 
  \lambda_{00} \left(\nu_1-\frac{1}{|h_{{\rm{rs}},1}|^2+|h_{{\rm{rs}},2}|^2} \right)^+
 +\lambda_{01} \left(\nu_1-\frac{1}{|h_{{\rm{rs}},1}|^2} \right)^+
 +\lambda_{10} \left(\nu_1-\frac{1}{|h_{{\rm{rs}},2}|^2} \right)^+ 
 =1.
\end{align*}
For case~(ii), it suffices to set  $\lambda_{01}=\lambda_{10}=0$ in case~(i).
Let $\lambda_{11}=1-\lambda_{00}=\lambda\in[0,1]$, and ${ \|\mathbf{h}_{{\rm{dr}}}\|^2= |h_{{\rm{dr}},1}|^2 + |h_{{\rm{dr}},2}|^2}, \ { \|\mathbf{h}_{{\rm{rs}}}\|^2= |h_{{\rm{rs}},1}|^2+|h_{{\rm{rs}},2}|^2}$. With this we get
\begin{align*}
\mathsf{C}^{\prime\prime}_\text{case~(ii)} &= \max_{\lambda\in[0,1]} \min\left\{
    \lambda \left.\log\left(1+\frac{\|\mathbf{h}_{{\rm{dr}}}\|^2}{\lambda}   \right)\right.
    ,
(1-\lambda) \left.\log\left(1+\frac{\|\mathbf{h}_{{\rm{rs}}}\|^2}{1-\lambda} \right)\right.
\right\}
\\& { \in \left[
 \frac{\log(1+\|\mathbf{h}_{{\rm{rs}}}\|^2)\log(1+\|\mathbf{h}_{{\rm{dr}}}\|^2)}{\log(1+\|\mathbf{h}_{{\rm{rs}}}\|^2)+\log(1+\|\mathbf{h}_{{\rm{dr}}}\|^2)},
\frac{\log(1+\|\mathbf{h}_{{\rm{rs}}}\|^2)\log(1+\|\mathbf{h}_{{\rm{dr}}}\|^2)}{\log(1+\|\mathbf{h}_{{\rm{rs}}}\|^2)+\log(1+\|\mathbf{h}_{{\rm{dr}}}\|^2)}+1
 \right],}
\end{align*}
where the optimal $\lambda$ is obtained by equating the two expressions within the min. 
}

\bibliographystyle{IEEEtran}
\bibliography{ITWBib}

\end{document}